\documentclass[11pt]{article}
\usepackage[dvips]{graphics}
\input{amssym.def}
\input{amssym.tex}

\usepackage{xcolor}

\title{\Large\bf 
Reciprocal tube  tight-binding approximation \\method 
for carbon nanotubes}
\author{Yuri A.  Antipov\\
yantipov@lsu.edu\\ 
Department of Mathematics, Louisiana State University\\
Baton Rouge LA 70803, USA}

\newcommand{\beq}{\begin{equation}}
\newcommand{\eeq}{\end{equation}}
\newcommand{\barr}{\begin{eqnarray}}
\newcommand{\earr}{\end{eqnarray}}

\newcommand{\Md}{\partial}

\newcommand{\ov}[1]{\overline{#1}}

\newcommand{\Ga}{\alpha}

\newcommand{\Ge}{\epsilon}

\newcommand{\Gf}{\phi}
\newcommand{\Gvf}{\varphi}
\newcommand{\Gg}{\gamma}

\newcommand{\Gk}{\kappa}

\newcommand{\Gl}{\lambda}
\newcommand{\Gn}{\eta}
\newcommand{\Gm}{\mu}

\newcommand{\Gt}{\theta}

\newcommand{\Gs}{\sigma}

\newcommand{\Go}{\omega}
\newcommand{\Gx}{\xi}

\newcommand{\GF}{\Phi}

\newcommand{\GP}{\Pi}

\newcommand{\CA}{{\cal A}}
\newcommand{\CB}{{\cal B}}

\newcommand{\CG}{{\cal G}}
\newcommand{\CH}{{\cal H}}

\newcommand{\CK}{{\cal K}}

\newcommand{\CM}{{\cal M}}

\newcommand{\CR}{{\cal R}}

\newcommand{\CT}{{\cal T}}
\newcommand{\CU}{{\cal U}}
\newcommand{\CV}{{\cal V}}

\def\Ba{{\bf a}}
\def\Bb{{\bf b}}

\def\Bk{{\bf k}}

\def\Br{{\bf r}}

\def\BC{{\bf C}}

\def\BT{{\bf T}}

\newcommand{\fr}{\frac}

\begin{document}
\maketitle

\begin{abstract}

The electronic structure  of a graphene sheet is altered when it is rolled up to form
a single-walled carbon nanotube (SWCNT), and the curvature effects for small radius nanotubes become significant.
In the paper, an analogue of the Bloch theory and tight-binding approximation for armchair, zigzag, and chiral SWCNTs is proposed.
The main features of the technique include the use of reciprocal nanotubes, translation-rotation transformations applied simultaneously   in the real and reciprocal
three-dimensional (3d) spaces, and construction of associated 2d  Brillouin zones and  Bloch double sums. 
Both sums have macroscopically large numbers of terms and built up by using two intersecting
helices. The method is efficient for all range nanotubes, narrow, intermediate-radius, and  megatubes. 
Electronic band structures calculations for sample chiral, armchair, and zigzag nanotubes are reported.

\end{abstract}


\setcounter{equation}{0}

\section{\label{S1}Introduction}

Although carbon nanotubes had been known since the middle of the last century, the active systematic study of electronic, optical and mechanical properties of 
nanotubes was initiated by the 1991 Iijima's report \cite{iij} that described high-resolution electron micrographs
of helicon carbon microtubes with the thinnest needle observed having 2.2 nm in diameter. It was pointed out that the helical structure of the needles was entirely different
from the screw dislocation model of conventional crystals in the sense that the needle crystals had a cylindrical lattice symmetry.  One of the most widely
applied techniques for describing the electronic  structure of SWCNTs is the zone-folding method \cite{sai98}, a  modification of
the tight-binding approximation approach developed for graphene \cite{wal}. Theoretical aspects of graphene and modern methods including the tight-binding approximation technique for describing the electronic structure  of single- and double-layer graphene were reviewed in \cite{net}.

The electronic band structure of SWCNTs analyzed by tight-binding approximation \cite{ham} predicted the existence of three classes of nanotubes, metallic, semiconducting with narrow band gaps, and 
 semiconducting with moderate band gaps.  Tight-binding folding of the 2d energy bands of graphene was applied  \cite{sai92a},  \cite{sai92b} to calculate the 
electronic  structure of SWCNTs. On the basis of the calculations it was predicted  \cite{sai92b} that approximately a third of SWCNTs  are a 1d metal, while the others
are 1d semiconductors. A similar procedure was implemented \cite{aji} to calculate the $\pi$-electron states of SWCNTs in magnetic field. The tight-binding approximation and folding the graphene energy bands were also employed \cite{sai93} for double-walled carbon commensurate and incommensurate nanotubes. It was concluded that
due to the symmetry, the interlayer interaction between the layers does not affect the metallic nature of the constituent.
The symmetry properties of chiral SWCNTs, the zone-folding technique, and the irreducible representations of symmetry groups associated with the chiral vector
were employed \cite{jis}, \cite{bar} to classify the electronic wave-functions and the phonon modes  at the Brillouin zone center.

Comparison \cite{bla} of two approaches for the electronic structure of SWCNTs, the graphene based tight-binding approximation and the method of pseudopotential  local density functional theory (DFT),
revealed that for small radius nanotubes in some cases the band gaps computed by the former theory deviated from those obtained by the second approach by more than 
50\%. Effects of curvature on the electronic structure of metallic SWCNTs were included \cite{kle} into the model by introducing a deformation tensor $D^{curv}=(d_{ij})$,
$i,j=1,2$, with the only one nonzero entry $d_{11}$ expressed through a transfer integral, the carbon-carbon bond length, and the nanotube radius.
 To take into account the curvature effects on the electronic
structure of SWCNTs, 
another approach,  a symmetry-adapted non-orthogonal tight-binding approximation scheme, was applied in \cite{pop}.
Electron-phonon interaction in an armchair nanotube $(5,5)$ and zigzag nanotubes $(5,0)$ and $(6,0)$
was analyzed in \cite{barn}. By using the Fr\"ohlich Hamiltonian it was found that  the zone-folding technique described well the tube $(5,5)$.  
In the case of the zigzag  nanotubes $(5,0)$ and $(6,0)$ the large
curvature made these tubes metallic with a large density of states at the Fermi energy and led to unusual
electron-phonon interactions, with the dominant coupling coming from the out-of-plane phonon modes. A DFT method for the analysis of torsional deformations 
of structures with helical symmetry was introduced in \cite{yu}. This method was recently adapted and  numerically
tested \cite{aga}  to  the Schr\"odinger
equation in helical coordinates.

The goal of this work was to further advance the tight-binding approximation technique  for SWCNTs and, particularly, for narrow and intermediate-radius nanotubes when
the obvious graphene-based approximation is not efficient.  
The method is based on a modification of the Bloch theory for armchair, zigzag and chiral nanotubes. 
Two features make it distinct from the methods used in the literature on nanotubes.
First, it introduces the concept of a reciprocal nanotube described in reciprocal cylindrical coordinates and employs, as a unit cell of the nanotube and its reciprocal,  a hexagon and parallelogram, respectively,
rolled around the tubes. Both tubes, the real nanotube and its reciprocal, are recovered by means of rotation-translation transformations of the unit cells in the real and reciprocal 3d spaces. The second characteristics  of the method is the way how the Bloch sum for tight-binding approximation is built up. 
The Bloch sum for chiral, armchair and zigzag  nanotubes  is designed as a double sum along two intersecting helices having macroscopically large numbers of atoms,
where the repeated terms are counted only once. The double sum has two wave numbers, and its structure naturally leads to a rectangular first Brillouin zone (a square in the armchair and zigzag cases).
This feature distinguishes the method from the zone-folding
technique  \cite{sai98} and the technique that employs the helical structure of SWCNTs \cite{whi}, \cite{zha}, and \cite{yu}. In the latter method, the Bloch sums are also
double. However, only one sum has a macroscopically large number of terms, while the number of terms in the second sum  is defined by the chirality numbers $n$ and $m$ and for narrow and 
intermediate-radius nanotubes is small. In both techniques,  \cite{sai98} and  \cite{whi}, the associated wave-vector is a 1d-vector, and the first Brillouin zone is a segment.

In Section \ref{S2}, we describe the structure and symmetry properties of  chiral,  armchair  and zigzag SWCNTs characterized by the chiral pairs $(n,m)$ $(1\le m\le n-1)$,
$(n,n)$, and $(n,0)$, respectively.
We introduce and characterize the associated reciprocal tubes, the rotation-translation transformations, and
the corresponding space groups in the real and reciprocal spaces and construct  a 2d first  Brillouin zone.
In Section \ref{S3}, based on the simultaneous
use of rotation-translation transformations,  $\CT^\pm_j$ and $\tilde\CT_j^\pm$, applied to the real nanotube and its reciprocal tube, respectively, 
we state and prove an analogue of the Bloch theorem
for 2d cylindrical quasi-hexagonal lattices with a curved parallelogram of periods. 
 We  prove that   $\psi(\CT_j^\pm\Br; \Bk)=e^{\pm ij  \Gk^\pm c_\pm}\psi(\Br;\Bk)$, where $\psi(\Br,\Bk)$ is the Bloch state, 
 $c_\pm$  are explicitly expressed through the chiral integers $n$ and $m$, 
$j=\pm 1,\pm 2, \ldots$,  $(\Gk^+,\Gk^-)\in \CB(n,m)$, and $\CB(n,m)$ is a first Brillouin rectangular   zone.
Also, we show that
$\psi(\Br;\Bk)=e^{i\Bk\cdot\Br}u(\Br;\Bk)$, where $u(\Br;\Bk)$ is  invariant with respect 
to the rotation-translation transformations applied simultaneously in the real and reciprocal spaces,
$u(\CT_j^\pm\Br; \tilde\CT_j^\pm\Bk)=u(\Br;\Bk)$.
In  Section \ref{S4}, we describe a tight-binding approximation scheme for $2p_z$-orbitals with $z$ orthogonal to the nanotube surface, 
derive the secular equation, and write down the Hamiltonian and overlap matrices associated with the first and second nearest-neighbor tight-binding approximations.
By establishing relations between the cylindrical coordinates associated the nanotube axis and the local spherical coordinates associated with the atoms employed by the scheme
we compute the Hamiltonian and overlap integrals. Finally, we report some numerical results on the electronic band structure of sample chiral, armchair, and zigzag nanotubes.

\setcounter{equation}{0}

\section{\label{S2} Chiral, armchair, and zigzag  nanotubes and their reciprocals}

\subsection{Chiral nanotube}

 \begin{figure}[t]
\centerline{
\scalebox{0.4}{\includegraphics{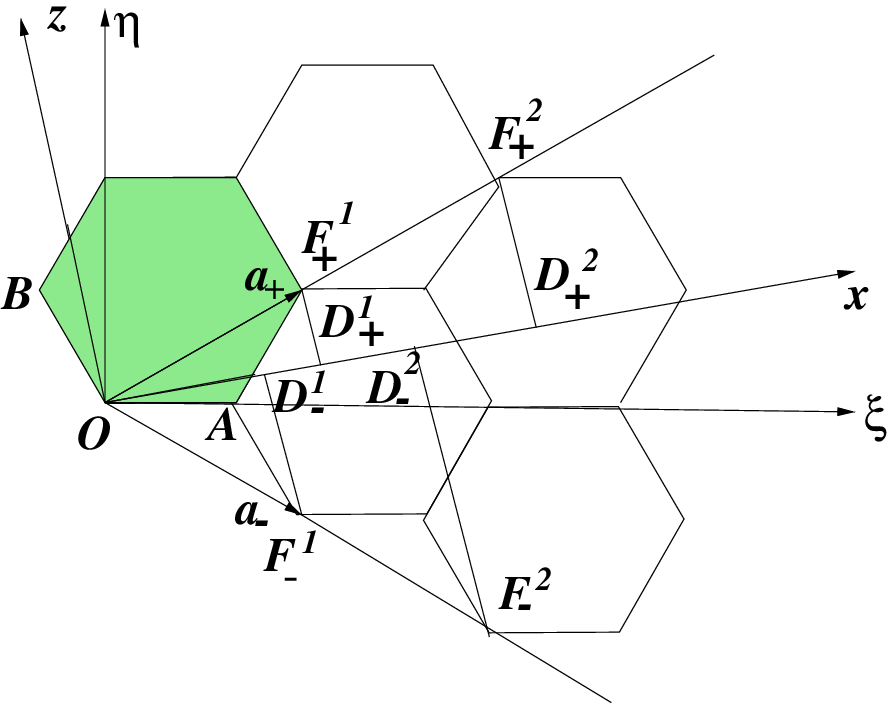}}
}
\caption{(a): Graphene sheet and the chiral and translation  vectors $\BC_h$ and $\BT$. (b): Projections of the vectors $j\Ba_\pm$ ($j=1,2$) on the $x$ and $z$-axes.}
\label{fig1}
\end{figure}

Let $(x,y,z)$ and $(r,\Gt,z)$ be Cartesian and cylindrical coordinates, respectively, with  the origin at $O_t$.
Consider a graphene sheet lying on the plane $\GP=\{|x|<\infty, y=r_t, |z|<\infty\}$ (Fig. \ref{fig1}) and characterized by the chiral vector $\BC_h=(n,m)=n\Ba_++m\Ba_-$,
where $n$ and $m$ are integers, $0<m<n$, $\Ba_\pm$ are the vectors
\beq
\Ba_+=\fr{a}{2}\langle \sqrt{3},1\rangle,\quad \Ba_-=\fr{a}{2}\langle \sqrt{3},-1\rangle,
\label{2.1}
\eeq
 in the frame $\Gx O \eta$, and $a=2.46$\AA  $\,$ is the lattice constant of graphene. 
The angles between the vector $\BC_h$ directed along the $x$-axis and the vectors $\Ba_+$ and  $\Ba_-$ are
\beq
\Gt_+=\cos^{-1}\fr{2n+m}{2\sqrt{n^2+m^2+nm}},
\quad
\Gt_-=\cos^{-1}\fr{n+2m}{2\sqrt{n^2+m^2+nm}}.
\label{3.3}
\eeq
The chiral vector $\BC_h$ is complemented by the translation vector $\BT$ \cite{sai98}
orthogonal to $\BC_h$, directing along the  $z$-axis and defined by
\beq
\BT=(t_1,t_2)=t_1 \Ba_+ +t_2 \Ba_-,
\quad
t_1=\fr{n+2m}{d_R}, \quad t_2=-\fr{2n+m}{d_R}.
\label{3.5}
\eeq
The number $d_R$ is expressed through the greatest common divisor $d$ of the integers $n+2m$ and $2n+m$ as follows.
If $n-m$ is a multiple of $3d$, then $d_R=3d$. Otherwise, $d_R=d$.

By rolling up the graphene sheet defined by the two vectors $\BC_h$  and $\BT$ and lying in the plane $\GP=\{y=r_t\}$ around the $z$-axis passing through the origin $O_t$
we construct a chiral nanotube $\CA=\CA(n,m)$. The circumference and radius $r_t$ of a cross-section of the tube orthogonal to its axis are
\beq
|\BC_h|=a\sqrt{n^2+m^2+nm}, 
\quad
r_t=\fr{|\BC_h|}{2\pi},
\label{3.6}
\eeq
while the nanotube  length is
\beq
|\BT|=\fr{\sqrt{3}a}{d_R}\sqrt{n^2+m^2+nm}.
\label{3.7}
\eeq
The nanotube designed can be extended along the $z$-axis to have the length $q|\BT|$,
$q$ is a positive real number.  The number of cells in the whole nanotube, $N$, is  assumed to be macroscopically large and multiple of $2m$, $N=2m L$.

Referring to the unrolled nanotube formed by the vectors $\BC_h$ and $\BT$ we define the projections
$OD_\pm^j$ and $F_\pm^jD_\pm^j$ of the vectors $j\Ba_\pm$
on the $x$- and $z$-axes, respectively (Fig. \ref{fig1}). They are
\beq
OD_\pm^j=ja\cos\Gt_\pm, \quad 
F_\pm^jD_\pm^j=\pm ja\sin\Gt_\pm, \quad
j=1,2,\ldots.
\label{3.8}
\eeq 
When the sheet is rolled up and the nanotube is formed, the projections $OD_+^1$ and $OD_-^1$ on the $x$-axis become circular arcs and
\beq
r_t\Ga_\pm=a\cos\Gt_\pm,
\label{3.9}
\eeq
where $\Ga_\pm$ are the angles of rotation of the points $D_\pm^1$ around the tube axis  when the point $O=(0,r_t,0)$ is fixed. This and formulas
(\ref{3.3}) and (\ref{3.9}) yield the angles of rotation
\beq
\Ga_+=\fr{\pi(2n+m)}{n^2+m^2+nm},\quad
\Ga_-=\fr{\pi(n+2m)}{n^2+m^2+nm}.
\label{3.10}
\eeq
The lattice vectors $\Ba_+$ and $\Ba_-$ become elliptical arcs, and their chords have the lengths (Table 1)
\beq
 a^\circ_\pm=\sqrt{a^2\sin^2\Gt_\pm+4r_t^2\sin^2\fr{\Ga_\pm}{2}}.
\label{3.11}
\eeq 

\vspace{5mm}

Table 1. The parameters $a^\circ_\pm/a$ for some chiral nanotubes.

\vspace{2mm}
\begin{tabular}{|c|c|c|c|c|c|}
\hline
\textrm{$\BC_h$ }&
\textrm{(4,2)} &
\textrm{(6,1)} &
\textrm{(6,5)} &
\textrm{(7,4)} &
\textrm{(8,3)} \\
\hline
 $a^\circ_+/a$                    &   0.9540        & 0.9635     & 0.9887  & 0.9867          & 0.9854      \\
\hline
 $a^\circ_-/a$                   &  0.9811         &  0.9947   &  0.9911      & 0.9936     &   0.9957    \\
 \hline
\end{tabular}

\vspace{2mm}

Introduce next the characteristic vectors of the chiral nanotube $\CA(n,m)$
\beq
\hat \Ba_+=\left(
\begin{array}{c}
-r_t\sin\Ga_+\\
r_t\cos\Ga_+\\
c_+\\
\end{array}
\right),\quad 
\hat \Ba_-=\left(
\begin{array}{c}
-r_t\sin\Ga_-\\
r_t\cos\Ga_-\\
-c_-\\
\end{array}
\right)
\label{3.12}
\eeq
where $\pm c_\pm$ are the projections of the vectors $\Ba_\pm$ on the nanotube axis and
\beq
c_+=a\sin\Gt_+=\fr{\sqrt{3}ma}{2\sqrt{n^2+m^2+nm}},
\quad
c_-=a\sin\Gt_-=\fr{\sqrt{3}na}{2\sqrt{n^2+m^2+nm}}.
\label{3.13}
\eeq

To describe the symmetry properties of the nanotube, we employ the following  rotation-translation transformations:
\beq
\CT_j^\pm\Br=\left(
\begin{array}{c}
-r_t\sin(j\Ga_\pm+\Gt) \\
r_t\cos(j\Ga_\pm+\Gt) \\
z\pm jc_\pm\\ 
\end{array}
\right), \quad j=\pm 1, \pm 2, \ldots,
\quad 
\Br=\left(
\begin{array}{c}
-r_t\sin\Gt\\
r_t\cos\Gt\\
z\\
\end{array}
\right)\in\CU_\CA,
\label{3.16}
\eeq
with $\CT_0^\pm$ being the identity map.
It becomes evident that if $s=0,1,\ldots, m-1$ and $j=-L+s, -L+s+1,\ldots, L+s-1$, then
the  transformations $\CM^{-+}_{s,j}=\CT^-_s\CT^+_j$ 
map the unit cell 
to all hexagons of the nanotube $\CA(n,m)$, and
each hexagon in $\CA$ is uniquely assigned to a pair $(s,j)$ provided
$\CM^{-+}_{s,-L+s}\CU^{00}_\CA=\CM^{-+}_{s,L+s}\CU^{00}_\CA$.
The transformations $\CM^{-+}_{s,j}$ ($s=0,1,\ldots,n-1, j=-L+s, -L+s+1,\ldots, L+s$) generate 
a space group say, $\CG_\CA$, whose fundamental domain  is  the interior of the unit cell $\CU_\CA$ complemented by the polygon line $BOAF_+^1$  (the vertices $B$ and  $F_+^1$ are excluded). 

We remark that another one-to-one correspondence between the pair $(j,s)$ and the cell $U^{js}_\CA$
is obtained if the transformations  $\CM^{+-}_{j,s}=\CT^+_j\CT^-_s$ are applied.  Here, $j=0,1,\ldots,n-1$ and $s=-L'+j,-L'+j+1,\ldots,L'+j-1$.
In this case we assume that the number of cells $N$ is a multiple of $2n$, $N=2n L'$.

Aiming to describe a Brillouin zone and derive an analogue of the Bloch theorem for nanotubes, as in the theory of lattices with translational symmetry, we introduce a reciprocal $\Bk$-space, $\tilde R^3$, and an analogue of the reciprocal lattice, a reciprocal tube,
$\tilde \CA=\tilde\CA(n,m)$. It will be convenient to operate with reciprocal Cartesian and cylindrical coordinates $(\tilde x, \tilde y, \Gk)$ and $(\tilde r,\Gt, \Gk)$, respectively.
The reciprocal radius of the cross-section of the tube $\tilde \CA$ is $\tilde r=\pi/r_t$, and the tube is characterized by the reciprocal tube vectors
\beq
\tilde \Bb_+=\left(
\begin{array}{c}
-\fr{\pi}{r_t}\sin\Ga_+\\
\fr{\pi}{r_t}\cos\Ga_+\\
\fr{2\pi}{c_+}\\
\end{array}
\right),\quad 
\tilde \Bb_-=\left(
\begin{array}{c}
-\fr{\pi}{r_t}\sin\Ga_-\\
\fr{\pi}{r_t}\cos\Ga_-\\
-\fr{2\pi}{c_-}\\
\end{array}
\right).
\label{3.14}
\eeq 
Notice that, in contrast to the hexagonal lattice vectors, the real nanotube vectors $\hat \Ba_\pm$
and the reciprocal vectors $\tilde\Bb_\mp$ are not mutually orthogonal and possess the properties
\beq
\hat\Ba_\pm\cdot\tilde\Bb_\pm=3\pi, \quad 
\hat\Ba_\pm\cdot\tilde\Bb_\mp=\pi\cos(\Ga_+-\Ga_-)-2\pi\fr{c_\pm}{c_\mp}. 
\label{3.15}
\eeq 
We select the  hexagon of the unrolled nanotube $\CA$ with the vertices $O$ and $F_+^1$ (it is shadowed in Fig. \ref{fig1}) as the preimage of the  unit cell of the tube.
The unit cell of the nanotube, $\CU_\CA$, is this hexagon rolled around the $z$-axis  passing  through the point $O_t$. If the integers $n$ and $m$ satisfy the condition
$0<m<n$, then all the sides of the unit cell $\CU_\CA$ are elliptical arcs.

\begin{figure}[t]
\centerline{
\scalebox{0.4}{\includegraphics{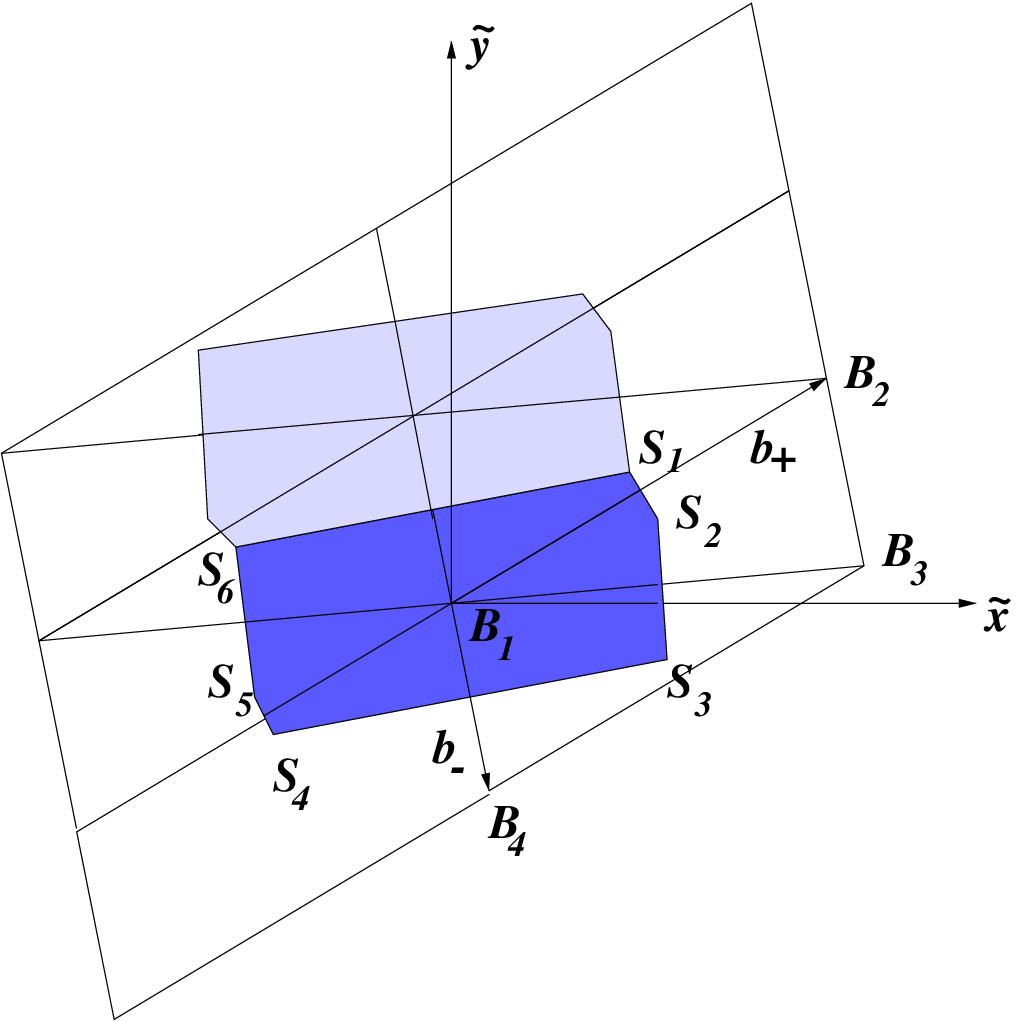}}
}
\caption{Chiral symmetry: the unrolled parallelogram of periods  $B_1B_2B_3B_4$ of the reciprocal tube $\tilde\CA$, 
 and the hexagon $S_1S_2\ldots S_6$.
}
\label{fig2}
\end{figure}

Consider now the reciprocal tube $\tilde\CA(n,m)$. We need to determine its curvilinear parallelogram of periods,  select a unit cell, and introduce an analogue of the transformations 
 $\CT_j^\pm$ in the reciprocal space $\tilde R^3$. 
 To do so, first, consider the reciprocal lattice characterized by two vectors $\Bb_\pm=\langle b_\pm', b_\pm''\rangle$ lying in the reciprocal plane $\tilde R^2$,
 with the
numbers $b_\pm'$ and $b_\pm''$ to be fixed.
 Denote the parallelogram of periods of the reciprocal lattice by  $B_1 B_2B_3 B_4$  (Fig. \ref{fig2}).
  The  flat reciprocal lattice  itself is described  by
 \beq
 \tilde \Br(l_1,l_2)=l_1\Bb_+ +l_2\Bb_-, 
 \label{3.17}
 \eeq
 where $l_1$ and $l_2$ are integers. The reciprocal tube is formed by rolling a portion of the reciprocal lattice around the $\Gk$-axis. 
 Based on the expressions (\ref{3.14}) of the reciprocal tube vectors, we find the second components of the vectors 
\beq
b_+''=\fr{2\pi}{c_+}, \quad  b_-''=-\fr{2\pi}{c_-}.
\label{3.18}
\eeq
Since $0<\Gt_+<\Gt_-<\fr{\pi}{3}$, we have $0<c_+<c_-$ and therefore $b_+''>|b_-''|$.
The first components of the vectors $\Bb_\pm$ are fixed by the condition
\beq
\Ga_\pm \tilde r_t=b_\pm'
\label{3.18'}
\eeq
and formulas  (\ref{3.6}) and (\ref{3.10}). We find
\beq
b_+'=\fr{2\pi^3(2n+m)}{a(n^2+m^2+nm)^{3/2}},\quad
b_-'=\fr{2\pi^3(n+2m)}{a(n^2+m^2+nm)^{3/2}}<b_+'.
\label{3.19}
\eeq

One of possible choices for a unit cell of the reciprocal tube $\tilde \CA$, $\tilde\CU_{\tilde\CA}=\tilde\CU^{00}_{\tilde\CA}$, is the image of the parallelogram $B_1B_2B_3B_4$ formed by the vectors $\Bb_+=B_1B_2$ and $\Bb_-=B_1B_4$
and rolled around the 
reciprocal tube $\tilde\CA$ (Fig. \ref{fig2}). By means of  compositions of the transformations $\tilde\CT_j^+$ and  $\tilde\CT_s^-$,  $\tilde\CM^{-+}_{s,j}=\tilde\CT^-_s\tilde\CT^+_j$,
defined in the reciprocal space $\tilde R^3$ as
\beq
\tilde\CT_j^\pm\Bk=\left(
\begin{array}{c}
-\fr{\pi}{r_t}\sin(j\Ga_\pm+\Gt) \\
\fr{\pi}{r_t}\cos(j\Ga_\pm+\Gt) \\
\Gk\pm\fr{2\pi j}{c_\pm}\\ 
\end{array}
\right), \quad 
\Bk=\left(
\begin{array}{c}
-\fr{\pi}{r_t}\sin\Gt\\
\fr{\pi}{r_t}\cos\Gt\\
\Gk\\
\end{array}
\right)\in\tilde\CU_{\tilde \CA},
\label{3.20}
\eeq
we establish a one-to-one correspondence between the pair $(s,j)$ ($s=0,1,\ldots m-1$, $j=-L+s,\ldots,L+s-1$)
and all  unit cells $\tilde\CU^{js}_{\tilde\CA}$
of the reciprocal tube. 

To construct an analogue of the first  Brillouin zone, consider four neighboring parallelogram-like cells of the reciprocal tube $\tilde \CA$  which share  the vertex $B_1$ (Fig. \ref{fig2}).
When unrolled, the four cells become parallelograms. The bisectors orthogonal to the lines joining the origin $B_1$ of the reciprocal coordinates $\tilde x, \tilde z$ 
with the six nearest lattice points form a polygon. It is the hexagon $S_1S_2\ldots S_6$ (Fig. \ref{fig2}), and its area and the area of the parallelogram $B_1B_2B_3B_4$ are the same.
The hexagonal cell is symmetric with respect to the origin $B_1$. When the hexagon is mapped back to the reciprocal tube $\tilde\CA$
by rolling it around its axis, the $\Gk$-axis passing through the origin of the reciprocal space, another reciprocal unit cell is formed.  
The domains of the transformations $\tilde\CT_j^+$ and $\tilde\CT_j^-$,  $\CK^+$ and $\CK^-$,
do not coincide. They are determined 
by dropping perpendiculars from the midpoints of the sides $S_1S_2$ and $S_4S_5$ for $\CK^+$ 
and $S_1S_6$ and $S_3S_4$ for $\CK^-$ on the reciprocal lattice axes. The domains 
\beq
\CK^\pm=\left\{\Bk \in\tilde\CA: -\fr{\pi}{c_\pm}\le\Gk^\pm\le \fr{\pi}{c_\pm}, -\fr{\Ga_\pm}{2} \le\Gt \le \fr{\Ga_\pm}{2}\right\}
\label{3.21}
\eeq
are two rectangles rolled around the surface of the reciprocal tube $\tilde \CA$.  Their unrolled preimages are
rectangles described in the reciprocal plane by $-\fr12 b_\pm'\le \tilde x\le \fr12 b_\pm'$, $-\fr{\pi}{c_\pm}\le\tilde y\le \fr{\pi}{c_\pm}$ with $b_\pm'$  and $c_\pm$ given by (\ref{3.19}) and (\ref{3.13}), respectively. 
The first  Brillouin zones $\CB(n,m)$ is the rectangle
 \beq
 \CB(n,m)=\left\{-\fr{\pi}{c_+}\le\Gk^+\le \fr{\pi}{c_+},  -\fr{\pi}{c_-}\le\Gk^-\le \fr{\pi}{c_-}\right\}.
 \label{3.21.0}
 \eeq

\subsection{Armchair nanotube and its reciprocal}

\begin{figure}[t]
\centerline{
\scalebox{0.4}{\includegraphics{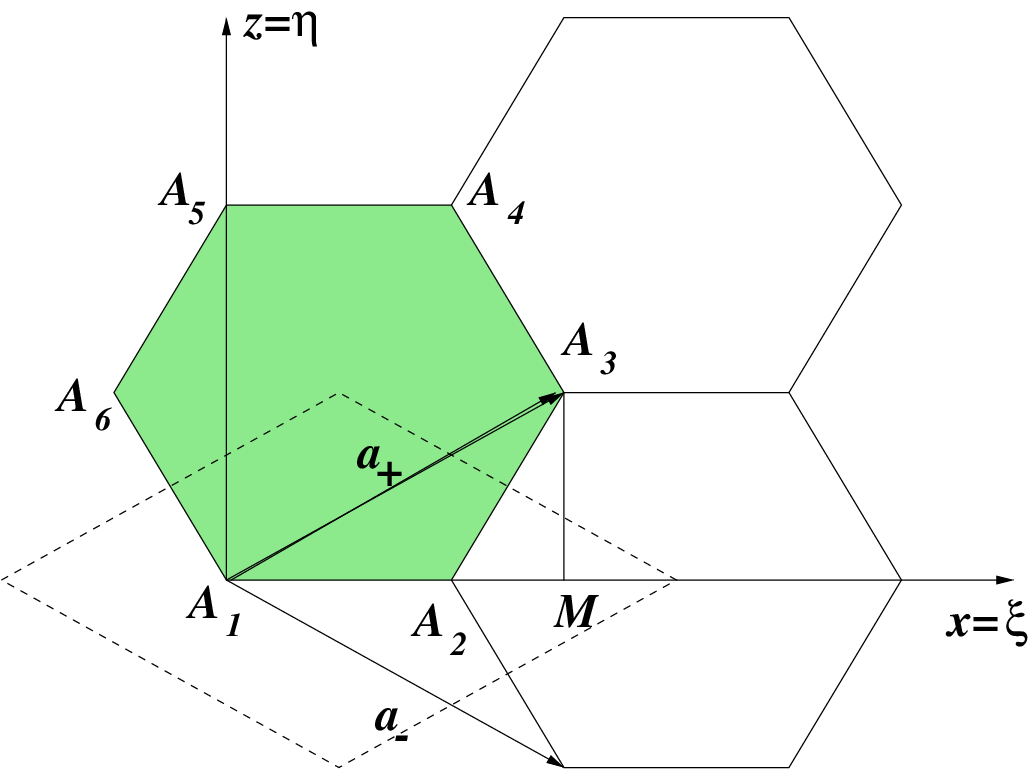}}
}
\caption{Hexagonal unit cell associated with an armchair nanotube.}
\label{fig3}
\end{figure}

An armchair nanotube,  $\CA(n,n)$, is characterized by the chiral vector $\BC_h=(n,n)$. 
Its radius $r_t$, the angles $\Gt_\pm$ and $\Ga_\pm$, and the parameters $c_\pm$ become (Fig. \ref{fig3}) 
\beq
r_t=\fr{na\sqrt{3}}{2\pi}, \quad \Gt_\pm=\fr{\pi}{6}, \quad \Ga=\Ga_\pm=\fr{\pi}{n},\quad c_\pm=\fr{a}{2}.
\label{3.22}
\eeq
The rolling of the graphene sheet transforms the vectors $\Ba_\pm$ into elliptical arcs whose chords have the length  (Table 2)
\beq
a^\circ=a\sqrt{\fr14+\fr{3n^2}{\pi^2}\sin^2\fr{\pi}{2n}}.
\label{3.23}
\eeq

\vspace{2mm}

Table 2. The parameter $a^\circ /a$ for some armchair nanotubes.

\vspace{2mm}
\begin{tabular}{|c|c|c|c|c|c|}
\hline
$\BC_h$ & (4,4) & (5,5) & (6,6) & (10,10) & (20,20) \\
\hline
$a^\circ /a$ &   0.9809 & 0.9877 &  0.9915 &    0.9969 & 0.9992  \\
\hline
\end{tabular}

 \begin{figure}[t]
\centerline{
\scalebox{0.35}{\includegraphics{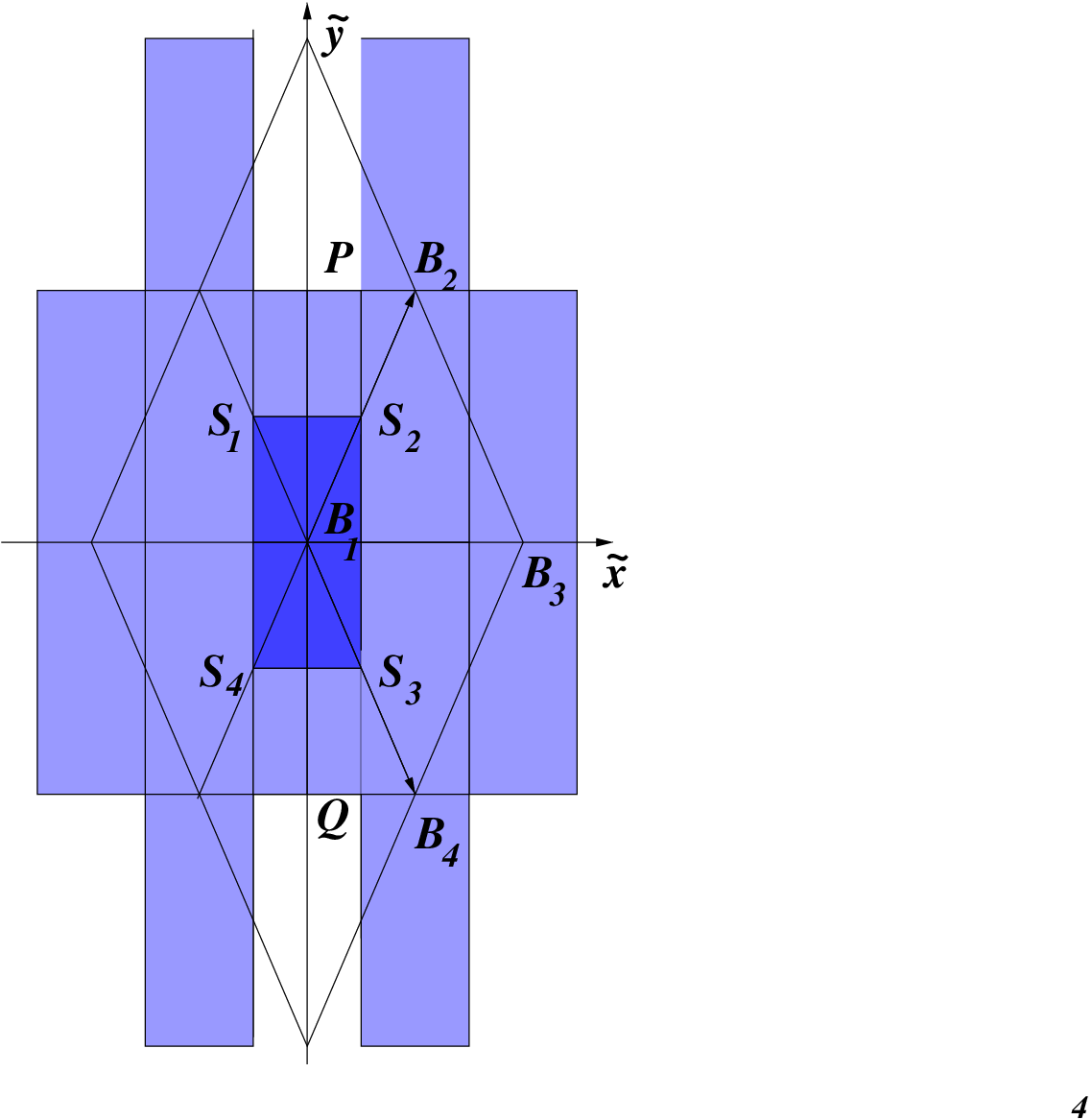}}
}
\caption{The domain $\CK=\CK^\pm$ of the transformations $\tilde\CT^\pm$ (the deep blue rectangle) of the reciprocal armchair nanotube $\tilde\CA$.}
\label{fig4}
\end{figure} 

The characteristic vectors of an armchair nanotube $\CA(n,n)$ and its reciprocal $\tilde \CA(n,n)$ have the form
\beq
\hat \Ba_\pm=\left(
\begin{array}{c}
-r_t\sin\Ga\\
r_t\cos\Ga\\
\pm\fr{a}{2}\\
\end{array}
\right),\quad 
\tilde \Bb_\pm=\left(
\begin{array}{c}
-\fr{\pi}{r_t}\sin\Ga\\
\fr{\pi}{r_t}\cos\Ga\\
\pm\fr{4\pi}{a}\\
\end{array}
\right),
\label{3.24}
\eeq
and the radius of the reciprocal tube is $\tilde r_t=\fr{2\pi^2}{a n\sqrt{3}}$.
The transformations $\tilde\CT^+$ and $\tilde\CT^-$  
share their domain  (Fig. \ref{fig4})
\beq
\CK^\pm=\left\{\Bk\in\tilde\CA: -\fr{2\pi}{a}\le\Gk^\pm\le\fr{2\pi}{a}, -\fr{\pi}{2n} \le\Gt \le\fr{\pi}{2n}\right\}.
\label{3.25}
\eeq
Therefore the first Brillouin zone becomes a square. It is given by  
\beq
\CB(n,n)=\left\{-\fr{2\pi}{a}\le\Gk^+\le\fr{2\pi}{a},  -\fr{2\pi}{a}\le\Gk^-\le\fr{2\pi}{a} 
\right\}.
\label{3.26}
\eeq

\subsection{Zigzag nanotube and  its reciprocal}

 \begin{figure}[t]
\centerline{
\scalebox{0.4}{\includegraphics{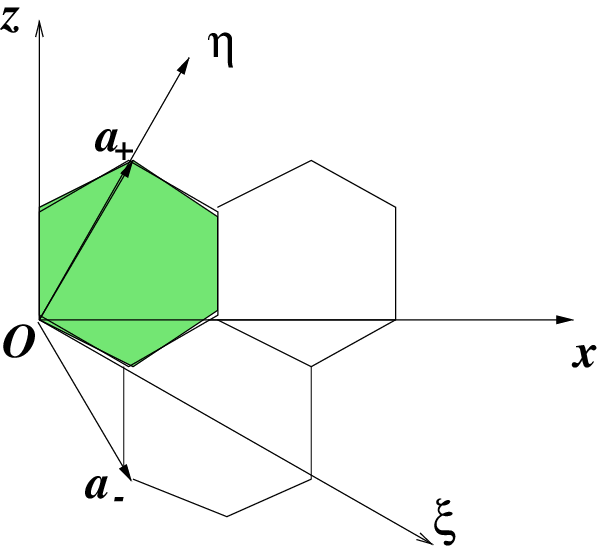}}
}
\caption{An unrolled zigzag nanotube}
\label{fig5}
\end{figure}

A  zigzag nanotube, $\CA(n,0)$, is characterized by the chiral vector $\BC_h=(n,0)$. The graphene sheet
associated with  the unrolled tube is placed on the plane $xOz$, $y=r_t$, and the frame  $\Gx O \eta$ forms an 
angle $\fr{\pi}{6}$ with the frame $x O z$ (Fig. \ref{fig5}).  For zigzag nanotubes, if the general formulas for a chiral nanotube $\BC_h(n,m)$
are employed, one of the associated transformations, $\CT^+_j$, becomes a pure rotation, and the tight-binding approximation method does not work.
Therefore, as the characteristic vectors for a graphene sheet, the preimage of the corresponding nanotube, we choose another set of vectors. They are
\beq
\Ba_+=\fr{a}{2}\langle 1, \sqrt{3}\rangle,\quad \Ba_-=\fr{a}{2}\langle  -1, \sqrt{3}\rangle.
\label{4.0}
\eeq
The radius $r_t$, the chiral angles $\Gt_\pm$, and the angles of rotations become
\beq
r_t=\fr{na}{2\pi}, \quad \Gt_+=\Gt_-=\fr{\pi}{3},\quad \Ga=\Ga_\pm=\fr{\pi}{n}.
\label{4.1}
\eeq
 The lattice vectors $\Ba_+$ and $\Ba_-$ become  elliptic arcs whose chords are of length  $a^\circ$ (Table 3),
 \beq
 a^\circ =a\sqrt{\fr{3}{4}+\fr{n^2}{\pi^2}\sin^2\fr{\pi}{2n}}.
 \label{4.3}
 \eeq

\vspace{5mm}

Table 3. The parameter $a^\circ/a$ for some zigzag nanotubes.

\begin{tabular}{|c|c|c|c|c|c|}
\hline
\textrm{$\BC_h$ }&
\textrm{(5,0)} &
\textrm{(6,0)} &
\textrm{(8,0)} &
\textrm{(10,0)} &
\textrm{(20,0)} \\
\hline
$a^\circ/a$                &   0.9959       &  0.9972   &   0.9984      &  0.9990   &   0.9997    \\
\hline
\end{tabular}

\vspace{2mm}

For a zigzag nanotube,  the characteristic vectors $\hat\Ba_\pm$ of the real zigzag  nanotube
$\CA(n,0)$ and the vectors $\tilde b_\pm$ of the reciprocal tube $\tilde\CA(n,0)$ become
\beq
\hat \Ba_\pm=\left(
\begin{array}{c}
-r_t\sin\Ga\\
r_t\cos\Ga\\
\pm\fr{\sqrt{3}a}{2}\\
\end{array}
\right),
\quad 
\tilde \Bb_\pm=\left(
\begin{array}{c}
-\fr{\pi}{r_t}\sin\Ga\\\
\fr{\pi}{r_t}\cos\Ga\\\
\pm\fr{4\pi}{\sqrt{3}a}\\
\end{array}
\right),
\label{4.5}
\eeq 
 while the rotation-translation transformations $\CT_j^\pm$ have the form
\beq
\CT_j^\pm\Br=\left(
\begin{array}{c}
-r_t\sin(j\Ga+\Gt) \\
r_t\cos(j\Ga+\Gt)  \\
z\pm\fr{\sqrt{3}aj}{2}\\ 
\end{array}
\right), \quad j=\pm 1, \pm 2, \ldots.
\label{4.7}
\eeq
The shadowed hexagon (Fig. \ref{fig5}) rolled around the nanotube is chosen as  a  unit cell $\CU_\CA=\CU^{00}_\CA$.
For zigzag nanotubes,  the number of cells in the tube is a multiple of $2n$, $N=2n L$.
The transformations   $\CM_{j,s}^{+-}$  ($j=0,1,\ldots,n-1$, $s=-L, -L+1,\ldots, L-1$) generate a space group $\CG_\CA$
and
map the unit cell  $\CU_\CA$ to all cells $\CU^{js}_\CA$. It is evident that there is a one-to-one correspondence between the pair
$(j,s)$ and  the cell $\CU^{js}_\CA$ provided $j=0,1,\ldots,n-1$, $s=-L, -L+1,\ldots, L-1$, and 
$\CM^{+-}_{j,-L}\CU^{00}_\CA=\CM^{+-}_{j,L}\CU^{00}_\CA$.

The reciprocal tube $\tilde \CA$ of the zigzag nanotube $\CA$ has radius $\tilde r_t=\fr{2\pi^2}{na}$ and the associated transformations 
 are
 \beq
\tilde\CT_j^\pm\Bk=\left(
\begin{array}{c}
-\tilde r_t\sin(j\Ga+\Gt) \\
\tilde r_t\cos(j\Ga+\Gt)  \\
\Gk\pm\fr{4\pi j}{\sqrt{3}a}\\ 
\end{array}
\right), \quad j=\pm 1, \pm 2, \ldots.
\label{4.9}
\eeq

 \begin{figure}[t]
\centerline{
\scalebox{0.4}{\includegraphics{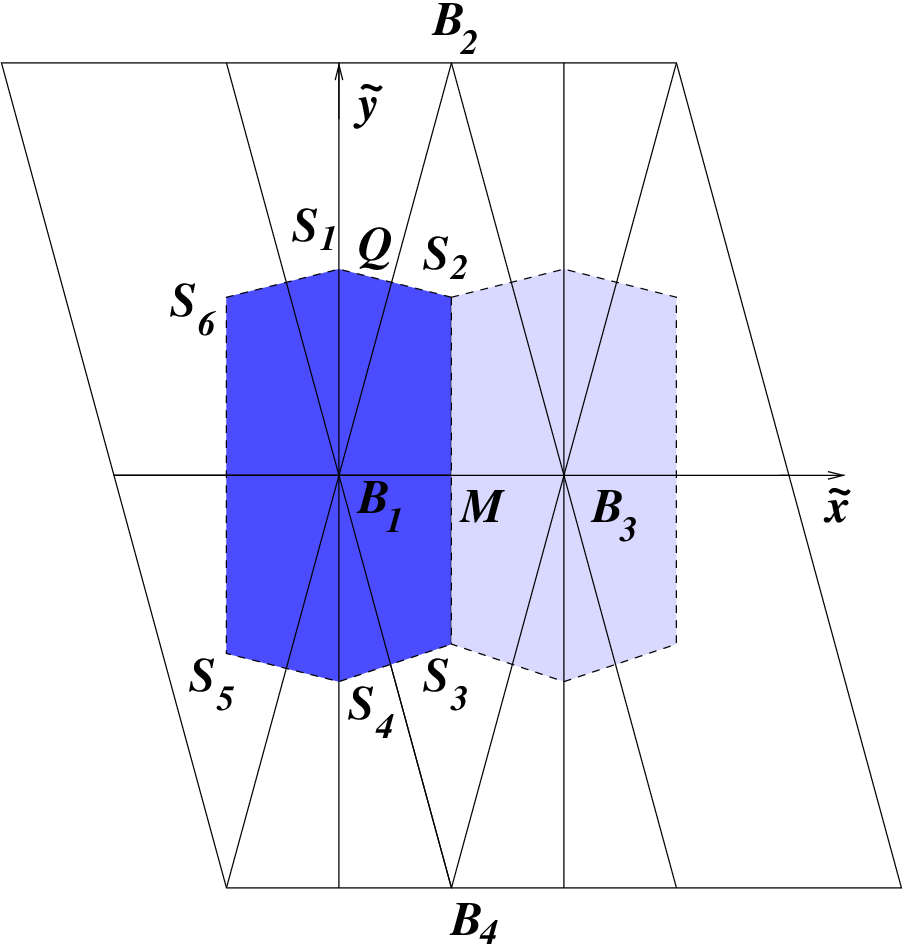}}
}
\caption{Zigzag symmetry: the parallelogram  of periods $B_1B_2B_3B_4$ of the unrolled
reciprocal tube $\tilde\CA$, 
and the hexagon $S_1S_2\ldots S_6$.}
\label{fig6}
\end{figure}

The parallelogram  of periods $B_1B_2B_3B_4$ of the unrolled reciprocal tube $\tilde\CA$ is defined by the vectors $\Bb_\pm=\langle b_\pm', b_\pm''\rangle$
of the reciprocal lattice in the plane $\tilde R^2$ (Fig. \ref{fig6}). 
Formulas  (\ref{4.5}) immediately yield  
\beq
n b_\pm'=\pi \tilde r_t, 
\quad
b_\pm''=\pm\fr{4\pi}{\sqrt{3}a},
\label{4.10}
\eeq 
 and therefore
 \beq
 \Bb_\pm=\langle \fr{2\pi^3}{n^2 a}, \pm\fr{4\pi}{\sqrt{3}a}\rangle.
 \label{4.11}
 \eeq
If we consider the  four parallelograms which share  the vertex  $B_1$ and  draw the bisectors orthogonal to the lines joining the origin
$B_1$ to the nearest six lattice points, we form a hexagon  $S_1S_2 \ldots S_6$ (Fig. \ref{fig6}). It has the same area as the parallelogram 
$B_1B_2B_3B_4$ and can also be chosen as a unit cell of the lattice (the deep blue hexagon in Fig. \ref{fig6}). When rolled around the axis of the reciprocal tube to be a part of it,
the transformed hexagon becomes a unit cell $\tilde\CU_{\tilde \CA}$ of the reciprocal tube $\tilde\CA$. 
On joining the midpoints of the sides $S_1S_2$,  $S_3S_4$,  $S_4S_5$, and $S_1S_6$, 
we draw a rectangle whose height and width are $2|b_-''|$ and $2b_-'$,
respectively. From this rectangle we determine the domains of the transformations $\CT^\pm$
\beq
\CK^\pm=\left\{\Bk\in\tilde\CA: -\fr{2\pi}{\sqrt{3}a}\le\Gk^\pm\le \fr{2\pi}{\sqrt{3}a}, -\fr{\pi}{2n} \le\Gt \le\fr{\pi}{2n}\right\}.
\label{4.12}
\eeq
As for an armchair  nanotube, a zigzag nanotube possesses a square Brillouin zone
\beq
\CB(n,0)=\left\{ -\fr{2\pi}{\sqrt{3}a}\le\Gk^+\le \fr{2\pi}{\sqrt{3}a}, -\fr{2\pi}{\sqrt{3}a}\le\Gk^-\le \fr{2\pi}{\sqrt{3}a}\right\}.
\label{4.14}
\eeq

The results of this section are concisely summarized by the following.

\vspace{.1in}

{\sl Theorem 2.1.}  The first Brillouin zone of a chiral nanotube $\CA(n,m)$ is the rectangle 
\beq
 \CB(n,m)=\left\{-\fr{\Gs_{nm}}{ma}\le\Gk^+\le \fr{\Gs_{nm}}{ma},  -\fr{\Gs_{nm}}{na}\le\Gk^-\le \fr{\Gs_{nm}}{na}\right\},
 \label{4.15}
 \eeq
where $a$ is the graphene lattice constant and
\beq
\Gs_{nm}=\fr{2\pi\sqrt{n^2+m^2+nm}}{\sqrt{3}}.
\label{4.16}
\eeq
For armchair $\CA(n,m)$  and zigzag $\CA(n,0)$ nanotubes  the first Brillouin zone  is a square defined by (\ref{3.26}) and (\ref{4.14}), respectively.

\setcounter{equation}{0}

\section{\label{S3} Analogue of the Bloch theorem for carbon nanotubes}

In this section we state and prove an analogue  of the graphene Bloch theorem  for  SWCNTs.

\vspace{.1in}

{\sl Theorem 3.1.} Let $\CA=\CA(n,m)$ be a SWCNT characterized by the chiral vector $\BC_h(n,m)$. Suppose the potential $\CV(\Br)$
of the  Hamiltonian
\beq
\CH=-\fr{\hbar^2}{2m}\nabla^2+\CV(\Br)
\label{5.1'}
\eeq
 of a single electron has the rotation-translation symmetry of the nanotube
\beq
\CV(\CT_j^\pm\Br)=\CV(\Br), \quad j=\pm 1,\pm 2, \ldots,
\label{5.2}
\eeq
where the transformations $\CT_j^\pm$ are defined on the nanotube  $\CA(n,m)$ by (\ref{3.16}),  $\Br\in \CU_\CA$, and $\CU_\CA$
is a unit curved hexagonal cell of the nanotube.
Then  any solution of the Schr\"odinger equation
\beq
\CH \psi(\Br;\Bk)=\Ge(\Bk) \psi(\Br;\Bk)
\label{5.2'}
\eeq
satisfies the relations
\beq
\psi(\CT_j^\pm\Br; \tilde\CT_j^\pm\Bk)=\exp\left\{\pm ij \left( \Gk^\pm c_\pm+\fr{2\pi z}{c_\pm}\right)\right\}\psi(\Br;\Bk).
\label{5.3}
\eeq
Also, if the transformations in the reciprocal space are skipped, then
\beq
\psi(\CT_j^\pm\Br; \Bk)=e^{\pm ij  \Gk^\pm c_\pm}\psi(\Br;\Bk), 
\quad
j=\pm 1,\pm 2, \ldots.
\label{5.3'}
\eeq
Here, $c_\pm$ are the real parameters (\ref{3.13}), $\Bk\in\CK^\pm$ is the  wave-vector (\ref{3.20}) with $\Gk=\Gk^\pm$, $\CK^\pm$ are the domains (\ref{3.21}), 
 and $\tilde \CT_j^\pm$ are  the transformations (\ref{3.20}) defined on the reciprocal tube. 

Moreover, if the function $\psi(\Br;\Bk)$ is written as 
\beq
\psi(\Br;\Bk)=e^{i\Bk\cdot\Br}u(\Br;\Bk),
\label{5.4}
\eeq
then the function $u(\Br;\Bk)$ is invariant with respect to the composition  of the transformations $ \CT_j^\pm$ and $\tilde \CT_j^\pm$,
\beq
u(\CT_j^\pm\Br; \tilde\CT_j^\pm\Bk)=u(\Br;\Bk), \quad 
j=\pm 1,\pm 2, \ldots.
\label{5.5}
\eeq

In particular, if the nanotube has the armchair symmetry, $\CA=\CA(n,n)$, then the relations (\ref{5.3}) have the form
\beq
\psi(\CT_j^\pm\Br; \tilde\CT_j^\pm\Bk)=\exp\left\{\pm ij \left(\fr{\Gk^\pm a}{2}+\fr{4\pi z}{a}\right)\right\}\psi(\Br;\Bk), 
\quad
\Bk\in\CK^\pm, \quad j=\pm 1,\pm 2, \ldots.
\label{5.6}
\eeq 
For a zigzag nanotube $\CA=\CA(n,0)$, the relations (\ref{5.3}) read
\beq
\psi(\CT_j^\pm\Br; \tilde\CT_j^\pm\Bk)=\exp\left\{\pm ij \left(\fr{\sqrt{3}\Gk^\pm a}{2}+\fr{4\pi z}{\sqrt{3}a}\right)\right\}\psi(\Br;\Bk),
\quad
 \Bk\in\CK^\pm, \quad
j=\pm 1,\pm 2, \ldots.
\label{5.7}
\eeq

\vspace{.1in}

{\sl Proof.} Let $f(\Br;\Bk)$ be a function of $\Br$ and $\Bk$, the radius-vectors of points on the real and reciprocal tubes $\CA$ and $\tilde\CA$, respectively. 
Define the rotation-translation transformations of the function $f(\Br;\Bk)$ by
\beq
 \CT_j^\pm f(\Br;\Bk)=f(\CT_j^\pm \Br;\Bk), 
\quad
\tilde \CT_j^\pm f(\Br;\Bk)=f(\Br;\tilde \CT_j^\pm\Bk).
\label{5.9}
\eeq
Denote by $\psi(\Br;\Bk)$ an eigenfunction of the transformations $\CT_j^\pm$ in the real space,
\beq
\CT_j^\pm\psi(\Br;\Bk)=\Gl_j^\pm(\Bk)\psi(\Br;\Bk).
\label{5.10}
\eeq
Simultaneously, it is  an eigenfunction of transformation $\tilde \CT_j^\pm$ in the reciprocal space,
\beq
\tilde \CT_j^\pm\psi(\Br;\Bk)=\Gm_j^\pm(\Br)\psi(\Br;\Bk),
\label{5.11}
\eeq
that is
\beq
\CT_j^\pm \tilde \CT_j^\pm\psi(\Br;\Bk)=\Gl_j^\pm(\Bk)\Gm_j^\pm(\Br)\psi(\Br;\Bk).
\label{5.12}
\eeq
As in the Bloch theory for flat lattices, the electron distribution $|\psi(\Br;\Bk)|$ has to have
the same rotation-translation periodic properties as the potential (\ref{5.2}). Therefore
\beq
|\CT_j^\pm \tilde \CT_j^\pm\psi(\Br;\Bk)|=|\psi(\Br;\Bk)|.
\label{5.13}
\eeq
This implies that the eigenvalues $\Gl_j^\pm(\Bk)$ and $\Gm_j^\pm(\Br)$ have the form
\beq
\Gl_j^\pm(\Bk)=e^{i\Gs_j^\pm (\Bk)}, \quad \Gm_j^\pm(\Br)=e^{i\Gn_j^\pm (\Br)},
\label{5.14}
\eeq
where the arguments $\Gs_j^\pm (\Bk)$ and $\Gn_j^\pm (\Br)$ are invariant to the rotations $\CR_j$.
Since the translations in the transformations $\CT_j^\pm$ and $\tilde \CT_j^\pm$ are applied along the $z$ and $\Gk$-axes
of the real and reciprocal spaces $R^3$ and $\tilde R^3$, respectively, and because of formulas (\ref{3.16}) and (\ref{3.20}) and the following relations for compositions
of the transformations $\CT_j^\pm$ and $\tilde \CT_j^\pm$:
$$
\CT_j^\pm\tilde \CT_j^\pm \CT_l^\pm\tilde \CT_l^\pm\psi(\Br;\Bk)=\psi(\CT_{j+l}^\pm\Br;\tilde \CT_{j+l}\Bk)
$$
\beq
=\Gl_j(\Bk)\Gl_l(\Bk)\Gm_j(\Br)\Gm_l(\Br)\psi(\Br;\Bk)
=\Gl_{j+l}(\Bk)\Gm_{j+l}(\Br)\psi(\Br;\Bk),
\label{5.15}
\eeq
we deduce
\beq
\Gs_j^\pm(\Bk)=\pm \Gk c_\pm j, \quad 
\Gn_j^\pm(\Br)=\pm \fr{2\pi zj}{c_\pm}.
\label{5.16}
\eeq
It is evident that the Hamiltonian operator commutes with the transformations $\CT_j^\pm$
defined on the real tube $\CA$. Also, it commutes with the transformations 
\beq
e^{-i\Gn_j^\pm(\Br)}\tilde \CT_j^\pm=\exp\left(\mp\fr{2\pi i zj}{c_\pm}\right)\tilde \CT_j^\pm,
\label{5.17}
\eeq
that is the commutator
\beq
\left[\exp\left(\mp\fr{2\pi i zj}{c_\pm}\right)\tilde \CT_j^\pm  \CT^\pm_j,\CH\right]=0.
\label{5.18}
\eeq
Therefore, without loss of generality, the eigenfunction chosen before may be considered as a solution  of the wave equation (\ref{5.2}). This implies that
 the relation 
\beq
\CT_j^\pm\tilde \CT_j^\pm \psi(\Br;\Bk)
=\exp\left\{\pm ij \left(\Gk^\pm c_\pm+\fr{2\pi z}{c_\pm}\right)\right\}\psi(\Br;\Bk)
\label{5.19}
\eeq
holds, and formula (\ref{5.3}) is proved. If the transformations $\tilde\CT_j$ are bypassed, then instead of formula (\ref{5.3}) we have the relation (\ref{5.3'}).

Introduce next a function $u(\Br,\Bk)$ such that $\psi(\Br;\Bk)=e^{i\Bk\cdot\Br}u(\Br;\Bk)$ and compute the scalar product
\beq
\CT_j^\pm\Br\cdot \tilde \CT_j^\pm\Bk
=\pi\cos(\Gt-\tau)+\Gk^\pm z\pm c_\pm \Gk^\pm j \pm\fr{2\pi zj}{c_\pm}+2\pi j^2.
\label{5.20}
\eeq
Since 
\beq
\Bk\cdot\Br=\pi\cos(\Gt-\tau)+\Gk^\pm z,
\label{5.21}
\eeq
we deduce
\beq
e^{i \CT_j^\pm\Br\cdot \tilde \CT_j^\pm\Bk}
=\exp\left\{i \left(\Bk\cdot\Br\pm c_\pm\Gk^\pm j \pm\fr{2\pi zj}{c_\pm}\right)\right\}. 
\label{5.22}
\eeq
By applying the transformations $\CT_j^\pm\tilde \CT_j^\pm$  to the function $\psi(\Br;\Bk)$
we obtain
$$
\psi(\CT_j^\pm\Br; \tilde\CT_j^\pm\Bk)=
e^{i\Bk\cdot\Br} 
e^{\pm ij(c_\pm \Gk^\pm+2\pi z/c_\pm)}
u(\CT_j^\pm\Br; \tilde\CT_j^\pm\Bk)
$$
\beq
=e^{\pm ij(c_\pm \Gk^\pm+2\pi z/c_\pm)}\psi(\Br; \Bk)
=e^{\pm ij(c_\pm \Gk^\pm+2\pi z/c_\pm)}e^{i\Bk^\pm\cdot\Br} 
u(\Br; \Bk),
\label{5.23}
\eeq
and the periodicity property  (\ref{5.5}) of the function $u(\Br; \Bk)$ becomes evident.

Formulas (\ref{5.6}) and (\ref{5.7}) immediately follow from the general relations (\ref{5.3}) if we put 
$c_\pm=\fr{a}{2}$ for the armchair symmetry and  $c_\pm=\fr{\sqrt{3}a}{2}$ for the zigzag case. The theorem is proved.

\vspace{.1in}

{\sl Remark 3.1.} If the transformations $\tilde\CT^\pm_j$ on the reciprocal tube are ignored, then the second part of Theorem 2 for SWCNTs,
formula (\ref{5.4}),  is invalid.
Indeed, since
\beq
e^{i\Bk\cdot\CT_j^\pm\Br}=e^{i[\pi\cos(j\Ga_\pm+\Gt-\tau)+\Gk^\pm z]}e^{\pm ij\Gk^\pm c_\pm},
\label{5.23'}
\eeq
there is no way to have a function $u(\Br;\Bk)$ satisfying the periodicity condition $u(\CT_j^\pm\Br; \Bk)=u(\Br;\Bk)$,
$j=\pm 1,\pm 2,\ldots$, such that $\psi(\Br;\Bk)=e^{i\Bk\cdot\Br}u(\Br;\Bk)$.

\setcounter{equation}{0}

\section{\label{S4}Tight-binding approximation for SWCNTs}

\subsection{Eigenvalue problem for the Hamiltonian}

Consider  the eigenvalue problem for the Hamiltonian 
\beq
\left(-\fr{\hbar^2}{2m}\nabla^2+\CV(\Br)\right) \psi(\Br;\Bk)=\Ge(\Bk) \psi(\Br;\Bk),
\label{5.27}
\eeq
 subject to the cyclic boundary conditions 
$$
\CM^{-+}_{s,-L}\psi(\Br;\Bk)=\CM^{-+}_{s,L}\psi(\Br;\Bk),
\quad
 \quad s=0,\ldots,m-1,
$$
\beq
\CM^{+-}_{-L',s}\psi(\Br;\Bk)=\CM^{+-}_{L',s}\psi(\Br;\Bk),
\quad
 \quad s=0,\ldots,n-1,
\label{5.28}
\eeq
when  $1 \le m\le n$, and
\beq
\CM^{-+}_{s,-L}\psi(\Br;\Bk)=\CM^{-+}_{s,L}\psi(\Br;\Bk),\quad \CM^{+-}_{-L,s}\psi(\Br;\Bk)=\CM^{+-}_{L,s}\psi(\Br;\Bk),
\quad
 \quad s=0,\ldots,n-1,
\label{5.29}
\eeq
for zigzag nanotubes, $m=0$. Here, as before, $\CM^{\pm \mp}_{s,j}=\CT_{s}^\pm \CT_{j}^\mp$.
 We note that  the cyclic boundary conditions (\ref{5.28}) and (\ref{5.29}) are  SWCNTs analogues of the Born-von Karman periodic boundary conditions
for flat lattices.

Analyze first  chiral and armchair nanotubes. On using Theorem 3.1 we simplify the boundary conditions (\ref{5.28}) as 
\beq
e^{-iL\Gk^+ c_+}\psi(\Br;\Bk)=e^{iL\Gk^+ c_+}\psi(\Br;\Bk), \quad e^{-iL'\Gk^- c_-}\psi(\Br;\Bk)=e^{iL'\Gk^- c_-}\psi(\Br;\Bk), 
\label{5.31}
\eeq
These conditions for chiral nanotubes  are satisfied if
$$
\Gk^+=\Gk^+_\nu=\fr{\pi\nu}{Lc_+}, \quad \nu=-L,-L+1,\ldots, L-1,
$$
\beq
\Gk^-=\Gk^-_\nu=\fr{\pi\nu}{Lc_-}, \quad \nu=-L',-L'+1,\ldots, L'-1,
\label{5.32}
\eeq
where $c_\pm$ are the parameters defined in (\ref{3.13}).  For armchair and zigzag nanotubes,  $L=L'$, $c_+=c_-$, and  $c_+=\fr{a}{2}$  for the former tube, while
$c_+=\fr{a\sqrt{3}}{2}$ for the latter.

\subsection{Hamiltonian and overlap matrices}

Let  $\Gvf_\mu(\Br)$ ($\mu=A,B$)  be an atomic orbital associated with one of the two carbon atoms in the unit hexagonal cell or, equivalently, the parallelogram of periods, and
orthogonal to the nanotube surface. When the tube is unrolled  and the surface becomes a graphene sheet, the functions $\Gvf_\mu$ are classified as $2p_z$-orbitals
with $z$ orthogonal to the graphene sheet. Consider a tight-binding Bloch function of a SWCNT $\CA(n,m)$. 
For armchair and chiral nanotubes $\CA(n,m)$, $1\le m\le n$,
it is taken to be
\beq
\GF_\mu(\Br; \Gk^+, \Gk^-)=\fr{1}{\sqrt{N}}\sum_{s=-L_n}^{L_n-1}{'}e^{i s\Gk^- c_-} \sum_{j=-L_m}^{L_m-1}{'} e^{-ij\Gk^+ c_+} 
 \CT^-_s\CT^+_j\Gvf_\mu(\Br), 
\label{6.1}
\eeq
where $N$ is the number of cells in the nanotube,  $L_n=\fr{N}{2n}$,  and $L_m=\fr{N}{2m}$.  For armchair and zigzag nanotubes,  $L_n=L_m=\fr{N}{2n}$. 
The two real numbers 
 $\Gk^\pm=\Gk^\pm_\nu$ are restricted to the first Brillouin zone described in Theorem 2.1.
The terms in the double sum are ordered according to the distance of the $(s,j)$-cell
from the original unit cell. The prime sign means that  none of the hexagons is listed twice. 

{\sl Remark  4.1.} We emphasize that the atomic functions, and therefore the functions  $\GF_\mu$, are defined in the 3d space, while the Bloch property stated in Theorem 3.1
is valid only on the surface of the nanotube. However, since the atomic functions rapidly decay when the distance from the atom is growing, we assume that these
 functions approximately satisfy the Bloch property outside the nanotube surface.

Following the tight-binding approximation scheme we derive the secular equation for two energy bands $\Ge =\Ge_\pm(\Gk^+,\Gk^-) $
\beq 
\left|\begin{array}{cc}
H_{AA}-\Ge S_{AA} & H_{AB}-\Ge S_{AB} \\
H_{BA}-\Ge  S_{BA} & H_{BB}-\Ge S_{BB} \\
\end{array}
\right|=0,
\label{6.4}
\eeq
where $H=(H_{\mu'\mu})$ and $S=(S_{\mu'\mu})$ ($\mu', \mu=1,2 $) are the Hamiltonian and overlap matrices whose entries are
\beq
H_{\mu' \mu}=\langle\GF_{\mu'}| \CH |\GF_\mu\rangle, \quad 
S_{\mu' \mu}=\langle\GF_{\mu'} | \GF_\mu\rangle.
\label{6.5}
\eeq
This brings us to the bonding $\pi$ and antibonding $\pi^*$ energy bands, $\Ge_+ $ and $\Ge_- $,
\beq 
\Ge_\pm =\fr{-Q\pm\sqrt{Q^2-4\det S \det H}}{2 \det S},
\label{6.6}
\eeq
where we introduced the notation
\beq
Q=S_{AB}H_{BA}+S_{BA}H_{AB}-S_{AA}H_{BB}-S_{BB}H_{AA}.
\label{6.7}
\eeq
The coefficients $S_{\mu'\mu}$ may be expressed through integrals of the atomic orbitals over the whole nanotube surface. In the case $1\le m\le n$ they are
\beq
S_{\mu'\mu}=\fr{1}{N}\sum_{s',s=-L_n}^{L_n-1}{'}e^{i (s-s')\Gk^- c_-}\sum_{j',j=-L_m}^{L_m-1} {'}
e^{-i(j-j')\Gk^+  c_+}
 \int\ov{\CT_{s'}^-\CT_{j'}^+\Gvf_{\mu'}(\Br)}
\CT_s^-\CT_j^+\Gvf_\mu(\Br)d\Br.
\label{6.8}
\eeq
On making the substitution $\Br'=\CT_{s}^-\CT_{j}^+\Br$ we  transform the coefficients to the form
 \beq
S_{\mu'\mu}=\fr{1}{N}\sum_{s',s=-L_n}^{L_n-1}{'}e^{i (s-s')\Gk^- c_-}\sum_{j',j=-L_m}^{L_m-1} {'}
e^{-i(j-j')\Gk^+  c_+}
 \int\ov{\Gvf_{\mu'}(\CT_{s'-s}^-\CT_{j'-j}^+\Br')}
\Gvf_\mu(\Br')d\Br'.
\label{6.9}
\eeq
We next replace the indices $s'-s=s''$ and $j'-j=j''$ and obtain
\beq
S_{\mu'\mu}=\sum_{s=-L_n}^{L_n-1}{'}e^{-is\Gk^- c_-}\sum_{j=-L_m}^{L_m-1} {'}
e^{i j \Gk^+ c_+}
 \int\ov{\Gvf_{\mu'}(\CT_s^-\CT_j^+\Br)}
\Gvf_\mu(\Br)d\Br.
\label{6.10}
\eeq
In the same way we transform the coefficients $H_{\mu'\mu}$
\beq
H_{\mu'\mu}=\sum_{s=-L_n}^{L_n-1}{'}e^{-is\Gk^- c_-}\sum_{j=-L_m}^{L_m-1} {'}
e^{i j \Gk^+ c_+}
\int\ov{\Gvf_{\mu'}(\CT_s^-\CT_j^+\Br)}
\CH \Gvf_\mu(\Br)d\Br.
\label{6.11}
\eeq
The entries of the Hamiltonian and overlap matrices associated with a zigzag nanotube ($m=0$) have the form
$$
H_{\mu'\mu}=\sum_{j=0}^{n-1}\sum_{s=-L_n}^{L_n-1} 
e^{-i s \Gk^- c_-}
\int\ov{\Gvf_{\mu'}(\CR_j\CT_s^-\Br)}
\CH \Gvf_\mu(\Br)d\Br,
$$
\beq
S_{\mu'\mu}=\sum_{j=0}^{n-1}\sum_{s=-L_n}^{L_n-1} 
e^{-i s \Gk^- c_-}
\int\ov{\Gvf_{\mu'}(\CR_j\CT_s^-\Br)}
 \Gvf_\mu(\Br)d\Br.
\label{6.12}
\eeq

\subsection{First and second nearest-neighbor tight-binding approximations}
 
 First we restrict ourselves to the assumption that the electrons hop to the nearest atoms only. 
 Then the diagonal $AA$ and $BB$ entries of the matrices $H$ and $S$ for any nanotube, $0\le m\le n$, are
 $$
 H_{\mu\mu}=\langle\Gvf_\mu(\Br)|\CH|\Gvf_\mu(\Br)\rangle, \quad 
 S_{\mu\mu}=\langle\Gvf_\mu(\Br)|\Gvf_\mu(\Br)\rangle.
 $$
The off-diagonal entries are sums of three integrals. If $1\le m\le n$, in the sums (\ref{6.10}) and (\ref{6.11}) 
we keep the $(s,j)$-terms with $(s,j)$ being $(0,0)$, $(0,1)$,  and $(1,0)$ for the entry $H_{AB}$ and
$(0,0)$, $(0,-1)$,  and $(-1,0)$ for the entry $H_{BA}$, while discard the others. We have
$$
H_{AB}=\langle\Gvf_A(\Br)|\CH|\Gvf_B(\Br)\rangle+
e^{i\Gk^+ c_+}\langle\Gvf_A(\CT_1^+\Br)|\CH|\Gvf_B(\Br)\rangle
+e^{-i\Gk^- c_-}\langle\Gvf_A(\CT_1^-\Br)|\CH|\Gvf_B(\Br)\rangle,
$$
\beq
H_{BA}=\langle\Gvf_B(\Br)|\CH|\Gvf_A(\Br)\rangle+
e^{-i\Gk^+ c_+}\langle\Gvf_B(\CT_{-1}^+\Br)|\CH|\Gvf_A(\Br)\rangle
+e^{i\Gk^- c_-}\langle\Gvf_B(\CT_{-1}^-\Br)|\CH|\Gvf_A(\Br)\rangle.
\label{6.13}
\eeq
The counterparts of these formulas for a zigzag nanotube read
$$
H_{AB}=\langle\Gvf_A(\Br)|\CH|\Gvf_B(\Br)\rangle+e^{ic_+(\Gk^+ -\Gk^- )}
\langle\Gvf_A(\CT_1^+\CT_1^- \Br)|\CH|\Gvf_B(\Br)\rangle
+e^{-i \Gk^- c_-}\langle\Gvf_A(\CT_{1}^-\Br)|\CH|\Gvf_B(\Br)\rangle,
$$
\beq
H_{BA}=\langle\Gvf_B(\Br)|\CH|\Gvf_A(\Br)\rangle
+e^{-ic_+(\Gk^+ -\Gk^-)}\langle\Gvf_B(\CT_{-1}^+\CT_{-1}^-\Br)|\CH|\Gvf_A(\Br)\rangle+
e^{i\Gk^- c_-}
\langle\Gvf_B(\CT^-_{-1}\Br)|\CH|\Gvf_A(\Br)\rangle.
\label{6.13'}
\eeq 
Notice that $\Gk^+$ and $\Gk^-$ vary within the first Brillouin zone defined for chiral, armchair, and zigzag nanotubes by  (\ref{4.15}), (\ref{3.26}), and (\ref{4.14}), respectively.  The entries of the overlap matrix $S$ are obtained from the corresponding 
entries of the matrix $H$ if the Hamiltonian is replaced by the identity operator.

 \begin{figure}[t]
\centerline{
\scalebox{0.3}{\includegraphics{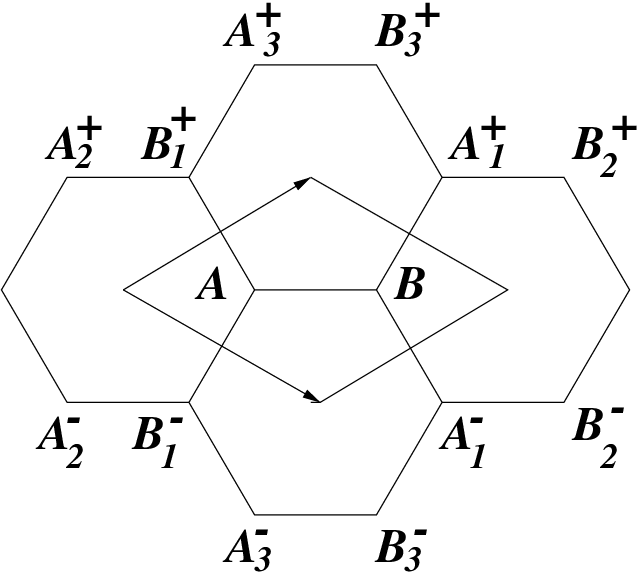}}
}
\caption{Unrolled four cells of chiral and armchair  nanotubes, a parallelogram of the periods,  and $A$- and $B$-sublattices.}
\label{fig7}
\end{figure}

 \begin{figure}[t]
\centerline{
\scalebox{0.3}{\includegraphics{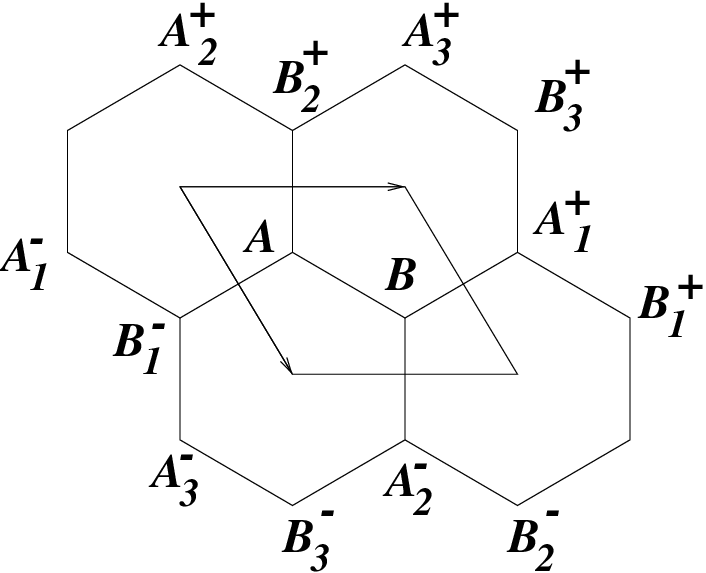}}
}
\caption{Unrolled four cells of a zigzag nanotube, a parallelogram of the periods,  and $A$- and $B$-sublattices.}
\label{fig8}
\end{figure}

In the second nearest-neighbor tight-binding approximation the off-diagonal entries are not changed, while the diagonal entries gain six 
more terms which reflect hopping between atom $\mu=A,B$ and the nearest atoms in the $\mu$-sublattice. 
These atoms are shown in Fig. \ref{fig7} for armchair  nanotubes
(their location is similar in the case $1\le m\le n)$ and in Fig. \ref{fig8} for zigzag nanotubes ($m=0$). 
For the entry $H_{AA}$ associated with a nanotube $\CA(n,m)$, $1\le m\le n$, we have
$$
H_{AA}=\langle\Gvf_A(\Br)|\CH|\Gvf_A(\Br)\rangle+\langle\Gvf_{A_1^+}(\Br)|\CH|\Gvf_A(\Br)\rangle
+\langle\Gvf_{A_1^-}(\Br)|\CH|\Gvf_A(\Br)\rangle+\langle\Gvf_{A_2^+}(\Br)|\CH|\Gvf_A(\Br)\rangle
$$
\beq
+\langle\Gvf_{A_2^-}(\Br)|\CH|\Gvf_A(\Br)\rangle+\langle\Gvf_{A_3^+}(\Br)|\CH|\Gvf_A(\Br)\rangle
+\langle\Gvf_{A_3^-}(\Br)|\CH|\Gvf_A(\Br)\rangle.
\label{6.14}
\eeq
The $(s,j)$-pairs in (\ref{6.11}) associated with the seven terms in the last relation are
$(0,0)$,  $(0,1)$,  $(1,0)$,  $(-1,0)$,  $(0,-1)$,  $(-1,1)$,  and $(1,-1)$, respectively. Therefore 
$$
H_{AA}=\langle\Gvf_A(\Br)|\CH|\Gvf_A(\Br)\rangle+e^{i\Gk^+ c_+}\langle\Gvf_{A}(\CT_1^+\Br)|\CH|\Gvf_A(\Br)\rangle
$$$$
+e^{-i\Gk^- c_-}\langle\Gvf_A(\CT_1^-\Br)|\CH|\Gvf_A(\Br)\rangle
+e^{i\Gk^- c_-}\langle\Gvf_{A}(\CT_{-1}^-\Br)|\CH|\Gvf_A(\Br)\rangle
$$
$$
+e^{-i\Gk^+ c_+}\langle\Gvf_{A}(\CT_{-1}^+\Br)|\CH|\Gvf_A(\Br)\rangle
+e^{i(\Gk^+ c_++\Gk^-c_-)}\langle\Gvf_{A}(\CT_1^+\CT_{-1}^-\Br)|\CH|\Gvf_A(\Br)\rangle
$$
\beq
+e^{-i(\Gk^+ c_++ \Gk^-c_-)}\langle\Gvf_{A}(\CT_1^-\CT_{-1}^+\Br)|\CH|\Gvf_A(\Br)\rangle.
\label{6.15}
\eeq  
Similarly,  the second diagonal entry is derived in the form
$$
H_{BB}=\langle\Gvf_B(\Br)|\CH|\Gvf_B(\Br)\rangle+e^{i\Gk^- c_-}\langle\Gvf_{B}(\CT_{-1}^-\Br)|\CH|\Gvf_B(\Br)\rangle
$$$$
+e^{-i\Gk^+ c_+}\langle\Gvf_B(\CT_{-1}^+\Br)|\CH|\Gvf_B(\Br)\rangle
+e^{i\Gk^+ c_+}\langle\Gvf_{B}(\CT_{1}^+\Br)|\CH|\Gvf_B(\Br)\rangle
$$
$$
+e^{-i\Gk^- c_-}\langle\Gvf_{B}(\CT_{1}^-\Br)|\CH|\Gvf_B(\Br)\rangle
+e^{i(\Gk^+c_++\Gk^- c_-)}\langle\Gvf_{B}(\CT_1^+\CT_{-1}^-\Br)|\CH|\Gvf_B(\Br)\rangle
$$
\beq
+e^{-i(\Gk^+c_+ +\Gk^-c_-)}\langle\Gvf_{B}(\CT_1^-\CT_{-1}^+\Br)|\CH|\Gvf_B(\Br)\rangle.
\label{6.16}
\eeq   
One can easily write down similar formulas for the $AA$ and $BB$ Hamiltonian integrals  in the second nearest-neighbor tight-binding approximation associated with  a zigzag nanotube. As in the first nearest-neighbor tight-binding approximation, the $AA$ and $BB$ entries of the overlap matrix $S$ are found from the 
above formulas for the corresponding entries  of the matrix $H$ if the  identity operator replaces the Hamiltonian.

\subsection{Computation of the overlap and Hamiltonian matrices }

For  numerical tests we choose the potential   $\CV_0(\Br')=-\fr{Ze^2}{r'}$, $Z=6$. 
The normalized $2p_{z'}$-hydrogenlike orbital $\Gf_A(\Br')$ in the local spherical coordinates $\Br'=(r',\Gf',\Gt')$ centered at the point $A$ with
$x=0$, $y=r_t$, $z=0$, and associated with the potential  $\CV_0(\Br')$
is given by   
\beq
\Gvf(\Br')=9\sqrt{\fr{3}{\pi}}r'e^{-3r'}\cos\Gt' \quad (\rm{atomic \; units, a.u.}).
\label{6.18}
\eeq
To evaluate the overlap and and Hamiltonian integrals, we notice that their integrands  are written in cylindrical system of coordinates, while the atomic functions
are in the spherical coordinates centered at the corresponding neighboring atoms. Consider the first nearest-neighbor tight-binding approximation.
The simplest integral of the overlap matrix is
\beq
S_{AA}=\int_0^\infty r dr\int_0^{2\pi}\ d\Gt\int_{-\infty}^\infty dz \tilde\Gvf(r,\Gt,z)
\label{6.19}
\eeq
where $\tilde\Gvf(r,\Gt,z)$ is the atomic function $\Gvf(\Br')$ given by (\ref{6.18}) and written in cylindrical coordinates. The cylindrical coordinates $(r,\Gt,z)$
\beq
x=-r\sin\Gt, \quad y=r\cos\Gt, \quad z=z,
\label{6.20}
\eeq
centered at (0,0,0) and
the local spherical coordinates $(r',\Gf',\Gt')$
\beq
x'=r'\cos\Gf'\sin\Gt', \quad y'=r'\sin\Gf'\sin\Gt', \quad z'=r'\cos\Gt'
\label{6.21}
\eeq
centered at the points $A$ and $B$ are linked by
\beq
x=x_\mu+x'\cos\Ga_\mu-z'\sin\Ga_\mu, \quad y=y_\mu+x' \sin\Ga_\mu+z'\cos\Ga_\mu, \quad z=z_\mu-y', \quad \mu=A,B,
\label{6.22}
\eeq
where 
$$
x_A=0, \quad y_A=r_t, \quad z_A=0, \quad x_B=-r_t\sin\Ga_0, \quad y_B=r_t\cos\Ga_0, \quad z_B=-c_0,
$$$$
\Ga_A=0, \quad \Ga_B=\Ga_0,
$$
\beq
\Ga_0=\fr{(n+m)\pi}{n^2+m^2+nm}, \quad c_0=\fr{(n-m)a}{2\sqrt{3(n^2+m^2+nm)}}.
\label{6.23}
\eeq
From here, the relations between 
the local $A$- and $B$-spherical coordinates and the tube cylindrical coordinates  $(r,\Gt,z)$
have the form
$$
r'=\sqrt{(x_\mu+r\sin\Gt)^2+(y_\mu-r\cos\Gt)^2+(z_\mu-z)^2}, 
$$
\beq
\cos\Gt'=\fr{1}{r'}[(x_\mu+r\sin\Gt)\sin\Ga_\mu-(y_\mu-r\cos\Gt)\cos\Ga_\mu],\quad \quad \sin\Gf'=\fr{z_\mu-z}{r'\sin\Gt'},
\label{6.24}
\eeq
and, inversely,
$$
r=\sqrt{(x_\mu-r'\cos\Gt'\sin\Ga_\mu+r'\cos\Gf'\sin\Gt'\cos\Ga_0)^2+(y_\mu+r'\cos\Gt'\cos\Ga_\mu+r'\cos\Gf'\sin\Gt'\sin\Ga_\mu)^2},
$$
$$
\cos\Gt=\fr{y_\mu+r'\cos\Gt'\cos\Ga_\mu+r'\cos\Gf'\sin\Gt'\sin\Ga_\mu}{r},
$$
\beq
z=z_\mu-r'\sin\Gf'\sin\Gt', \quad \mu=A,B.
\label{6.25}
\eeq
On computing  the Jacobian 
\beq
J(r',\Gf',\Gt')=\det\fr{\Md(r,\Gt,z)}{\Md(r',\Gf',\Gt')}
\label{6.26}
\eeq
 we find
\beq
r(r',\Gf',\Gt')|J(r',\Gf',\Gt')|=(r')^2\sin\Gt'
\label{6.27}
\eeq
and therefore 
\beq
S_{AA}=\int_0^\infty (r')^2 dr'\int_0^{2\pi}d\Gf'\int_0^\pi\sin\Gt'\Gvf^2(\Br') d\Gt' =1.
\label{6.28}
\eeq
Similar computations give us $S_{BB}=1$.

To evaluate the off-diagonal entries of the overlap matrix, we write down the radius-vectors of the atoms involved
\beq
\Br_{A_j}=\left(\begin{array}{c} x_{A_j}\\ y_{A_j}\\ z_{A_j}\\
\end{array}\right),\quad 
\Br_{B_j}=\left(\begin{array}{c} x_{B_j}\\ y_{B_j}\\ z_{B_j}\\
\end{array}\right), \quad j=1,2,3.
\label{6.29}
\eeq 
For chiral and armchair tubes these vectors are specified as
$$
\Br_{A_1}=\left(\begin{array}{c} 0 \\ r_t\\ 0\\\end{array}\right), 
\quad 
\Br_{A_2}=\CT_1^+\Br_{A_1}=\left(\begin{array}{c} -r_t\sin\Ga_+ \\r_t\cos\Ga_+\\  c_+\\ 
\end{array}\right), \quad  \Br_{A_3}=\CT_1^-\Br_{A_1}=\left(\begin{array}{c} -r_t\sin\Ga_- \\ r_t\cos\Ga_-\\ -c_-\\
\end{array}\right), 
$$
$$
\Br_{B_1}=\left(\begin{array}{c} -r_t\sin\Ga_0 \\ r_t\cos\Ga_0\\ -c_0\\\end{array}\right), 
\quad 
\Br_{B_2}=\CT_{-1}^+\Br_{B_1}=\left(\begin{array}{c} -r_t\sin(\Ga_0-\Ga_+) \\r_t\cos(\Ga_0-\Ga_+)\\  -c_0- c_+\\ 
\end{array}\right), 
$$
\beq
  \Br_{B_3}=\CT_{-1}^-\Br_{B_1}=\left(\begin{array}{c} -r_t\sin(\Ga_0-\Ga_-) \\ r_t\cos(\Ga_0-\Ga_-)\\ -c_0+c_-\\
\end{array}\right).
\label{6.30}
\eeq
The corresponding formulas for zigzag nanotubes have the form
$$
\Br_{A_1}=\left(\begin{array}{c} 0 \\ r_t\\ 0\\\end{array}\right), 
\quad 
\Br_{A_2}=\left(\begin{array}{c} -r_t\sin 2\Ga \\r_t\cos 2\Ga\\  0\\ 
\end{array}\right), \quad  \Br_{A_3}=\left(\begin{array}{c} -r_t\sin\Ga\\ r_t\cos\Ga \\ -\fr{\sqrt{3} a}{2}\\
\end{array}\right), 
$$
\beq
\Br_{B_1}=\left(\begin{array}{c} -r_t\sin\Ga \\ r_t\cos\Ga\\ -\fr{a}{2\sqrt{3}}\\\end{array}\right), 
\quad 
\Br_{B_2}=\CT_{-1}^+\CT_{-1}^-\Br_{B_1}=\left(\begin{array}{c}  r_t\sin\Ga \\r_t\cos\Ga\\ -\fr{a}{2\sqrt{3}} \\ 
\end{array}\right), \quad
  \Br_{B_3}=\CT_{-1}^-\Br_{B_1}=\left(\begin{array}{c} 0 \\ r_t \\ \fr{a}{\sqrt{3}}\\
\end{array}\right).
\label{6.30'}
\eeq

For any chiral nanotubes including armchair and zigzag tubes on establishing the links between the spherical coordinates centered at the left and right atoms in the integrals 
$\langle\Gvf_{A_j}(\Br)|\Gvf_B(\Br)\rangle$ and $\langle\Gvf_{B_j}(\Br)|\Gvf_A(\Br)\rangle$ we write 
$$
r'_{A_j}=\sqrt{(x_{B_1}-x_{A_j}+r'\cos\Gf'\sin\Gt')^2+(y_{B_1}-y_{A_j}+r'\cos\Gt')^2+(z_{B_1}-z_{A_j}-r'\sin\Gf'\sin\Gt')^2}
$$
\beq
\cos\Gt'_{A_j}=\fr{y_{B_1}-y_{A_j}+r'\cos\Gt'}{r'_{A_j}},
\label{6.31}
\eeq
and 
$$
r'_{B_j}=\sqrt{(x_{A_1}-x_{B_j}+r'\cos\Gf'\sin\Gt')^2+(y_{A_1}-y_{B_j}+r'\cos\Gt')^2+(z_{A_1}-z_{B_j}-r'\sin\Gf'\sin\Gt')^2}
$$
\beq
\cos\Gt'_{B_j}=\fr{y_{A_1}-y_{B_j}+r'\cos\Gt'}{r'_{B_j}},
\label{6.32}
\eeq
We now use these local spherical coordinates to express the off-diagonal entries of the matrix $S$ in terms of integrals that are easy to compute.
For chiral and armchair nanotubes, we find
$$
S_{AB}=S_{AB}^{(1)}+e^{i\Gk^+ c_+}S_{AB}^{(2)} +e^{-i\Gk^- c_-}S_{AB}^{(3)},
$$
\beq
S_{BA}=S_{BA}^{(1)}+e^{-i\Gk^+ c_+}S_{BA}^{(2)} +e^{i\Gk^- c_-}S_{BA}^{(3)},
\label{6.25.1}
\eeq
where
$$
S_{AB}^{(j)}=\int_0^\infty(r')^2dr'\int_0^{2\pi}d\Gf'\int_0^\pi\sin\Gt' \Gvf_{A_j}(\Br')\Gvf(\Br')d\Gt'
$$
\beq
S_{BA}^{(j)}=\int_0^\infty(r')^2dr'\int_0^{2\pi}d\Gf'\int_0^\pi\sin\Gt' \Gvf_{B_j}(\Br')\Gvf(\Br')d\Gt'
\label{6.26.1}
\eeq
$\Gvf(\Br')$ is the atomic function (\ref{6.18}), while the functions $\Gvf_{A_j}(\Br')$ and $\Gvf_{B_j}(\Br')$ are 
\beq
\Gvf_{\mu}(\Br')=9\sqrt{\fr{3}{\pi}}r'_\mu e^{-3r'_\mu}\cos\Gt'_\mu \quad \mu=A_j, B_j.
\label{6.27.1}
\eeq
The extension of the analysis to the entries of the Hamiltonian matrix presents only one new feature - the Hamiltonian operator originally written in the cylindrical system of coordinates 
centered at $(0,0,0)$ needs to be rewritten in the local spherical coordinates centered at the atoms involved.

Consider the nanotube  Hamiltonian
\beq
\CH=-\fr12\left(\fr{\Md^2}{\Md r^2}+\fr{1}{r}\fr{\Md}{\Md r}+\fr{1}{r^2}\fr{\Md^2}{\Md\Gt^2}+\fr{\Md^2}{\Md z^2}\right)+\CV(\Br)
\label{6.28.1} 
\eeq
with  the external potential chosen to be
\beq
\CV(\Br)=\sum_{\mu=A,B}\sum_{s=-L_n}^{L_n-1}\sum_{j=-L_m}^{L_m-1}{'}v^\mu_{sj}(\Br),
\label{6.29.1}
\eeq
where the atomic units are accepted and $(r,\Gt,z)$ is the cylindrical system of coordinates introduced in Section 1.
 As in the case of the overlap matrix, we transform the integrals in the Hamiltonian matrix originally written in the cylindrical system of coordinates to 
 integrals in the local spherical coordinates. For the diagonal entries we have    
 \beq
 H_{\mu\mu}=\int_0^\infty(r')^2dr'\int_0^{2\pi}d\Gf'\int_0^\pi\sin\Gt' \Gvf(\Br')\CH_\mu' \Gvf(\Br')d\Gt', \quad \mu=A,B.
\label{6.30.1}
\eeq
Similar to the off-diagonal entries of the overlap matrix we write for chiral and armchair nanotubes,
 $$
H_{AB}=H_{AB}^{(1)}+e^{i\Gk^+ c_+}H_{AB}^{(2)} +e^{-i\Gk^- c_-}H_{AB}^{(3)},
$$
\beq
H_{BA}=H_{BA}^{(1)}+e^{-i\Gk^+ c_+}H_{BA}^{(2)} +e^{i\Gk^- c_-}H_{BA}^{(3)},
\label{6.31.1}
\eeq
where
$$
H_{AB}^{(j)}=\int_0^\infty(r')^2dr'\int_0^{2\pi}d\Gf'\int_0^\pi\sin\Gt' \Gvf_{A_j}(\Br')\CH_B'\Gvf(\Br')d\Gt',
$$
\beq
H_{BA}^{(j)}=\int_0^\infty(r')^2dr'\int_0^{2\pi}d\Gf'\int_0^\pi\sin\Gt' \Gvf_{B_j}(\Br')\CH_A'\Gvf(\Br')d\Gt', \quad j=1,2,3.
\label{6.32.1}
\eeq
Here,
\beq
\CH_\mu'\Gvf(\Br')= -\fr12\left(\fr{\Md^2}{\Md r^2}+\fr{1}{r}\fr{\Md}{\Md r}+\fr{1}{r^2}\fr{\Md^2}{\Md\Gt^2}+\fr{\Md^2}{\Md z^2}\right)\Gvf(\Br')+v(\Br')\Gvf(\Br'),
\label{6.33}
\eeq
$v(\Br') =-\fr{Ze^2}{r'}$ if $r'\le  a/2$ and 0 otherwise, and
the derivatives are computed by the chain rule. The $r$-derivatives are 
$$
\fr{\Md \Gvf(\Br')}{\Md r}=\fr{\Md \Gvf(\Br')}{\Md r'}\fr{\Md r'}{\Md r}+\fr{\Md \Gvf(\Br')}{\Md \Gt'}\fr{\Md \Gt'}{\Md r},
$$
\beq
\fr{\Md^2 \Gvf(\Br')}{\Md r^2}=\fr{\Md^2 \Gvf(\Br')}{\Md r'^2}\left(\fr{\Md r'}{\Md r}\right)^2+\fr{\Md^2 \Gvf(\Br')}{\Md \Gt'^2}\left(\fr{\Md \Gt'}{\Md r}\right)^2
+\fr{2\Md^2  \Gvf(\Br')}{\Md r'\Md \Gt'}\fr{\Md r'}{\Md r}
\fr{\Md \Gt'}{\Md r}+\fr{\Md  \Gvf(\Br')}{\Md r'}\fr{\Md^2 r'}{\Md r^2}
+\fr{\Md  \Gvf(\Br')}{\Md \Gt'}\fr{\Md^2 \Gt'}{\Md r^2}.
\label{5.34}
\eeq
The second derivatives with respect to $\Gt$ and $z$ coincide with the second $r$-derivative if $r$ is replaced by $\Gt$ and $z$, respectively.
The derivatives of $r'$ and $\Gt'$ associated with $\CH'_A$ and $\CH'_B$ are computed from formulas (\ref{6.24}).
Their expressions  are long and omitted. For zigzag nanotubes, the factors $e^{\pm i\Gk^+ c_+}$
in formulas (\ref{6.25}) and (\ref{6.31}) need to be replaced by $e^{\pm ic_+(\Gk^+ -\Gk^-)}$, respectively.

\subsection{Numerical results}

\begin{figure}[t]
\centerline{
\scalebox{0.5}{\includegraphics{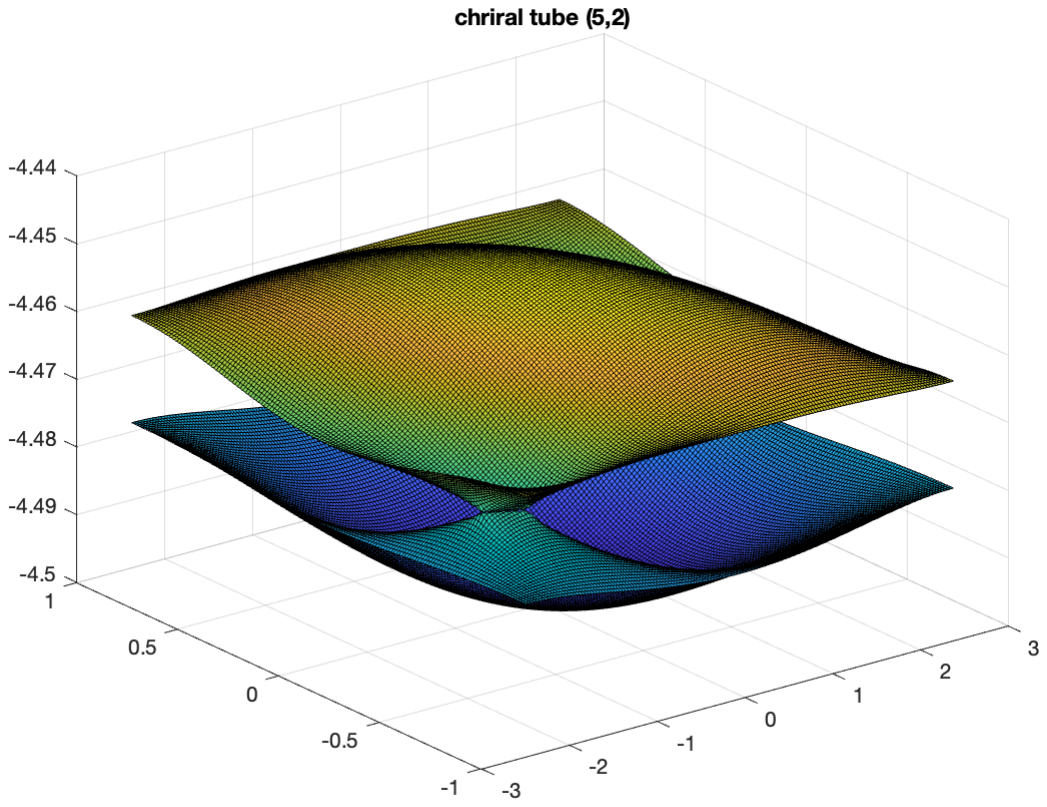}}
}
\caption{The 2d energy dispersion  relation in the first Brillouin zone for a chiral nanotube $\CA(5,2)$.}
\label{fig9}
\end{figure}

\begin{figure}[t]
\centerline{
\scalebox{0.5}{\includegraphics{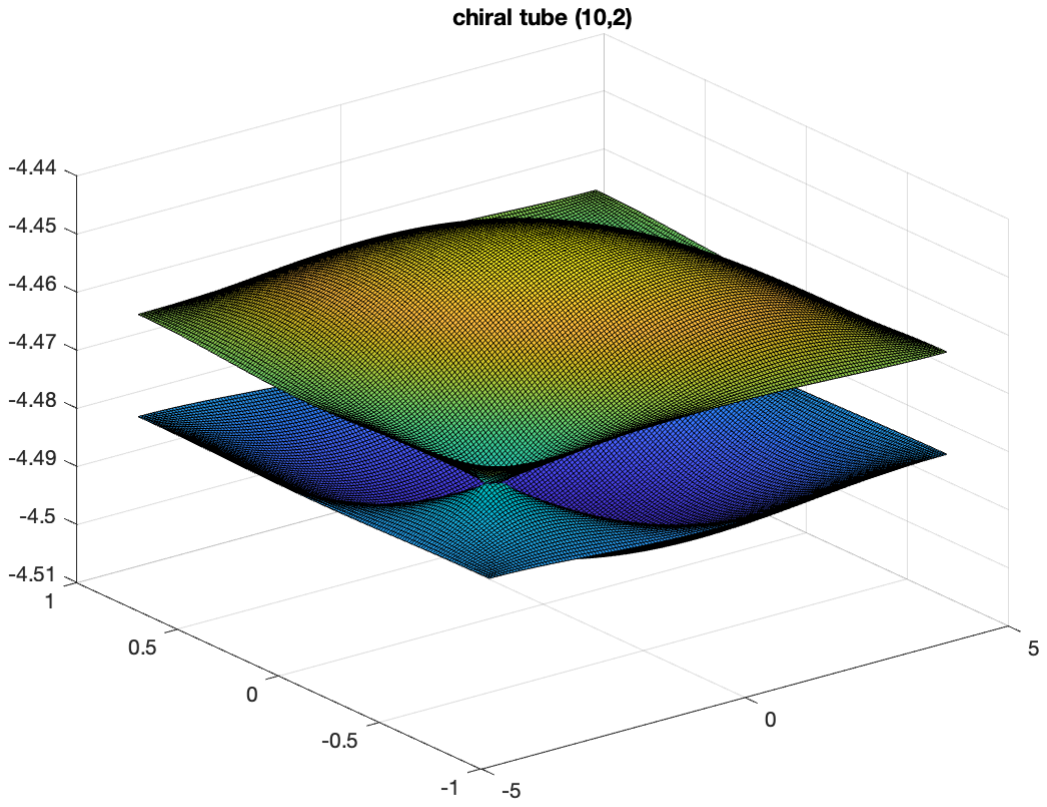}}
}
\caption{The 2d energy dispersion  relation in the first Brillouin zone for a chiral nanotube $\CA(10,2)$.}
\label{fig10}
\end{figure}

\begin{figure}[t]
\centerline{
\scalebox{0.5}{\includegraphics{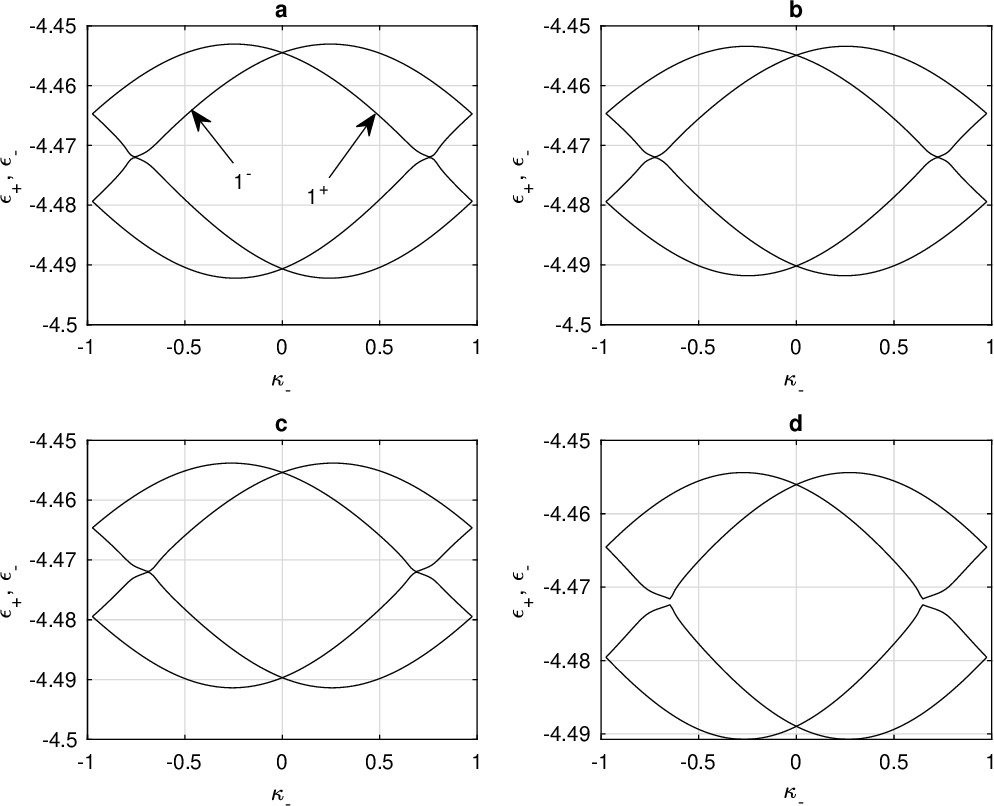}}
}
\caption{The energy bands $\epsilon_+$ and  $\epsilon_-$  versus $\Gk_-$ 
for a chiral nanotube $\CA(5,2)$ when (a) $\Gk_+=-0.57\pi/c_+$ (curve $1^-$) and $\Gk_+=0.57\pi/c_+$ (curve $1^+$), (b) $\Gk_+=-0.59\pi/c_+$ and  $\Gk_+=0.59\pi/c_+$,
(c)  $\Gk_+=-0.61\pi/c_+$ and  $\Gk_+=0.61\pi/c_+$, and  (d) $\Gk_+=-0.64\pi/c_+$ and  $\Gk_+=0.64\pi/c_+$.}
\label{fig11}
\end{figure}

\begin{figure}[t]
\centerline{
\scalebox{0.5}{\includegraphics{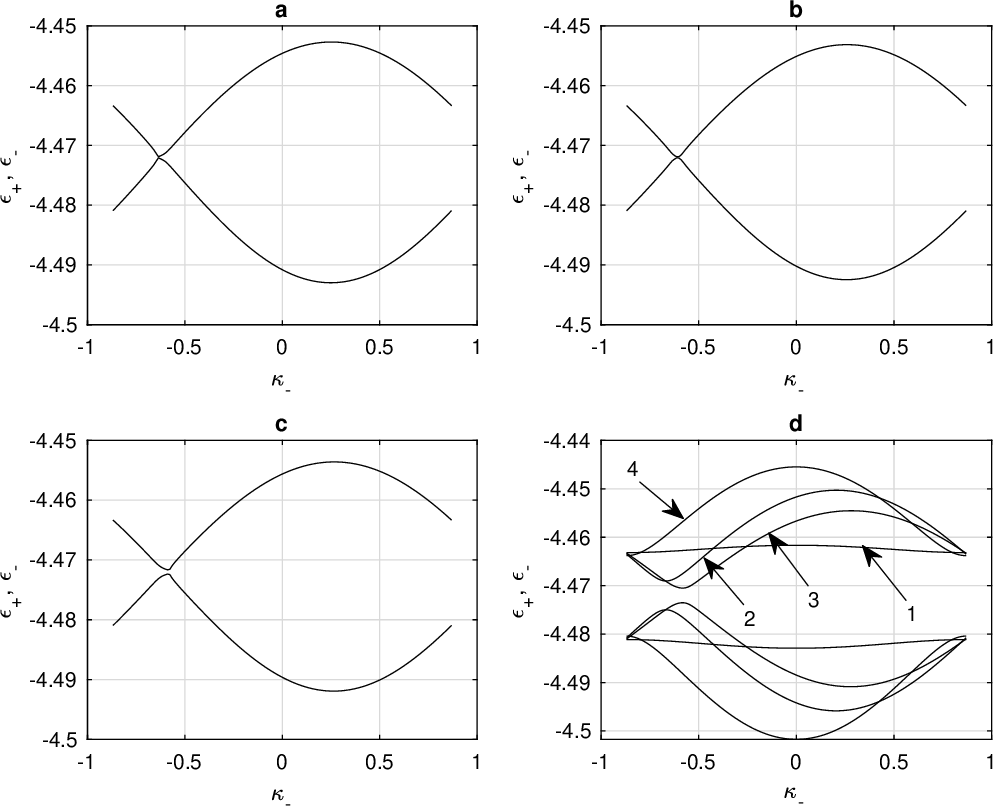}}
}
\caption{The energy bands $\epsilon_+$ and  $\epsilon_-$  versus $\Gk_-$ 
for a chiral nanotube $\CA(10,2)$ when (a) $\Gk_+=-0.62\pi/c_+$,   (b)  $\Gk_+=-0.64\pi/c_+$,  and (c)  $\Gk_+=-0.66\pi/c_+$.  In case (d)   curves 1, 2, 3, and 4 correspond to
$\Gk_+=-\pi/c_+$,  $-0.7\pi/c_+$, $-0.5\pi/c_+$, and $0$.} 
\label{fig12}
\end{figure}

\begin{figure}[t]
\centerline{
\scalebox{0.5}{\includegraphics{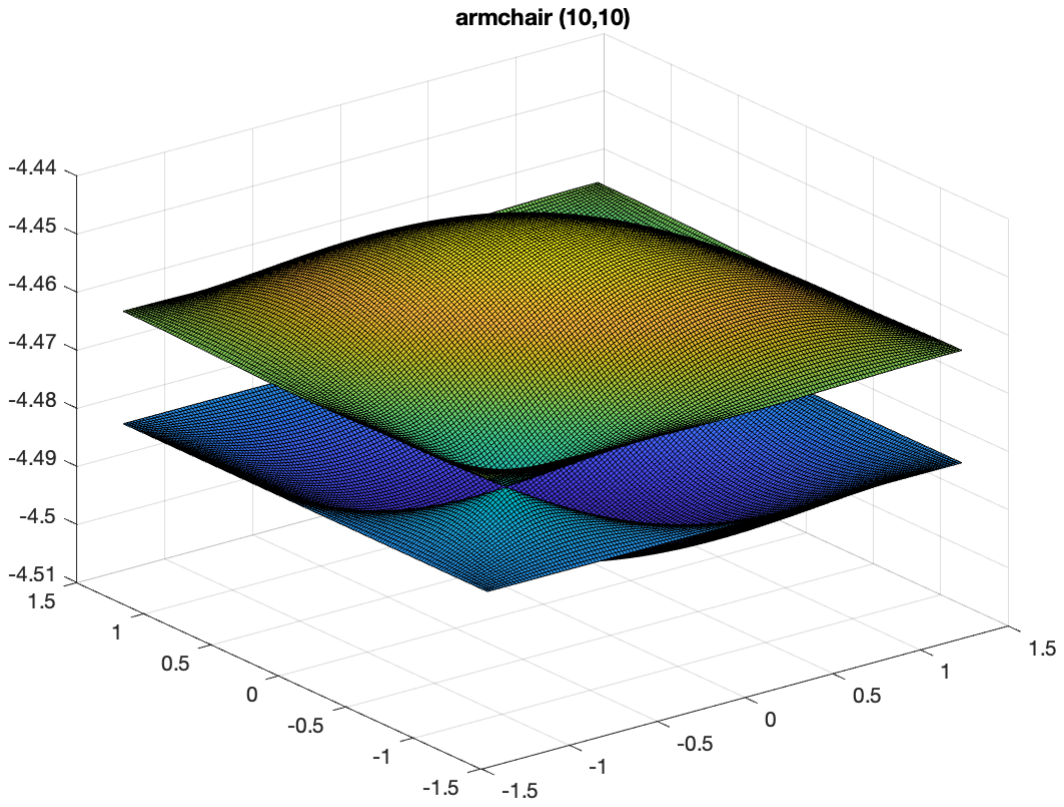}}
}
\caption{The 2d energy dispersion  relation in the first Brillouin zone for an armchair tube $\CA(10,10)$.} 
\label{fig13}
\end{figure}

\begin{figure}[t]
\centerline{
\scalebox{0.5}{\includegraphics{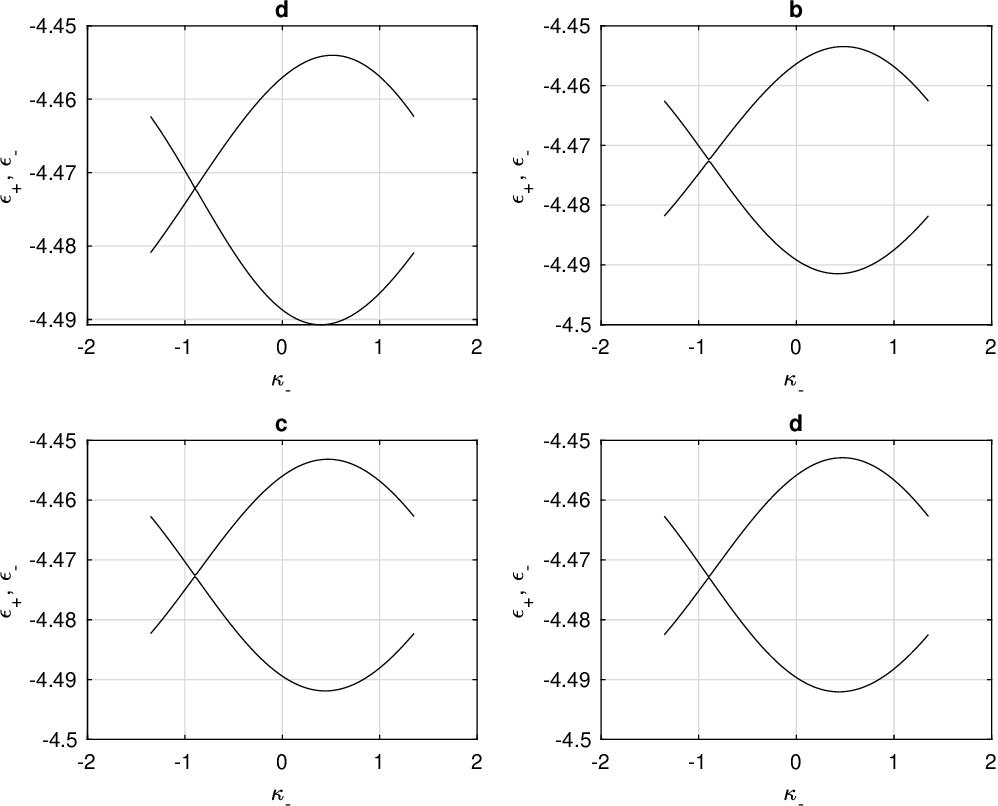}}
}
\caption{The energy bands $\epsilon_+$ and  $\epsilon_-$  versus $\Gk_-$ 
for armchair nanotubes $\CA(n,n)$  for $\Gk_\pm=-0.664\pi/c_+$ when  $\Ge_+-\Ge_-$ attains its minimum and  $n=5$ (a), 
$n=7$ (b), $n=10$ (c), and $n=15$ (d).} 
\label{fig14}
\end{figure}

\begin{figure}[t]
\centerline{
\scalebox{0.5}{\includegraphics{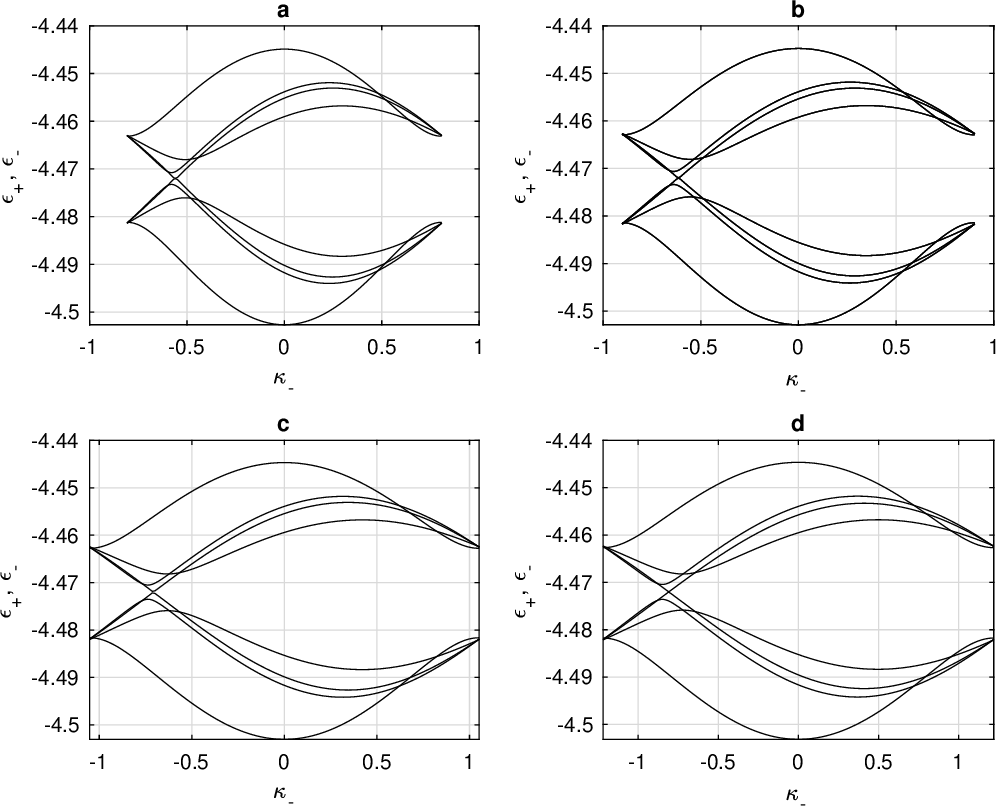}}
}
\caption{The energy bands $\epsilon_+$ and  $\epsilon_-$  versus $\Gk_-$ 
for chiral nanotubes $\CA(15,m)$ with  $m=1$ (a), 
$m=4$ (b), $m=8$ (c), and $m=12$ (d). The Dirac-like cone is observed for $\Gk_+=-0.65 \pi/c_+$ (a),  $-0.655 \pi/c_+$ (b), $-0.66 \pi/c_+$ (c),
$-0.664 \pi/c_+$ (d). The other curves (a-d) are for $\Gk_+=-0.4 \pi/c_+$, $\Gk_+=-0.6 \pi/c_+$, and $\Gk_+=-0.8 \pi/c_+$.} 
\label{fig15}
\end{figure}

\begin{figure}[t]
\centerline{
\scalebox{0.5}{\includegraphics{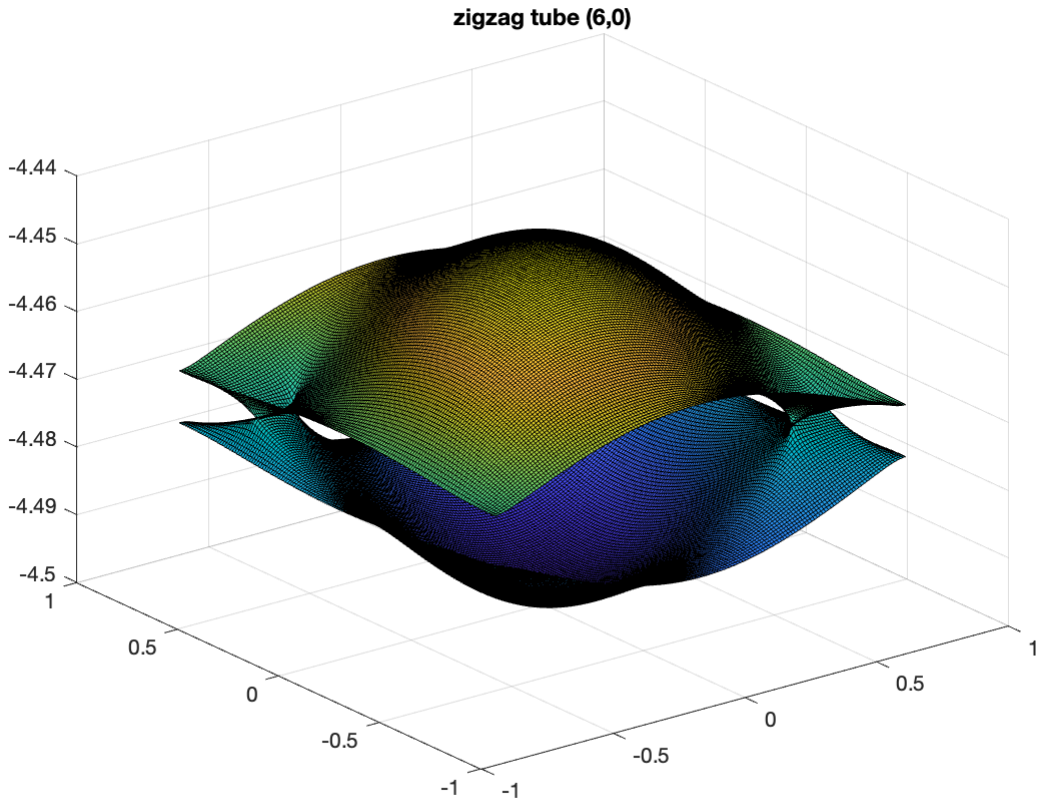}}
}
\caption{The energy dispersion  relation in the first Brillouin zone for a zigzag tube $\CA(6,0)$.} 
\label{fig16}
\end{figure}

\begin{figure}[t]
\centerline{
\scalebox{0.5}{\includegraphics{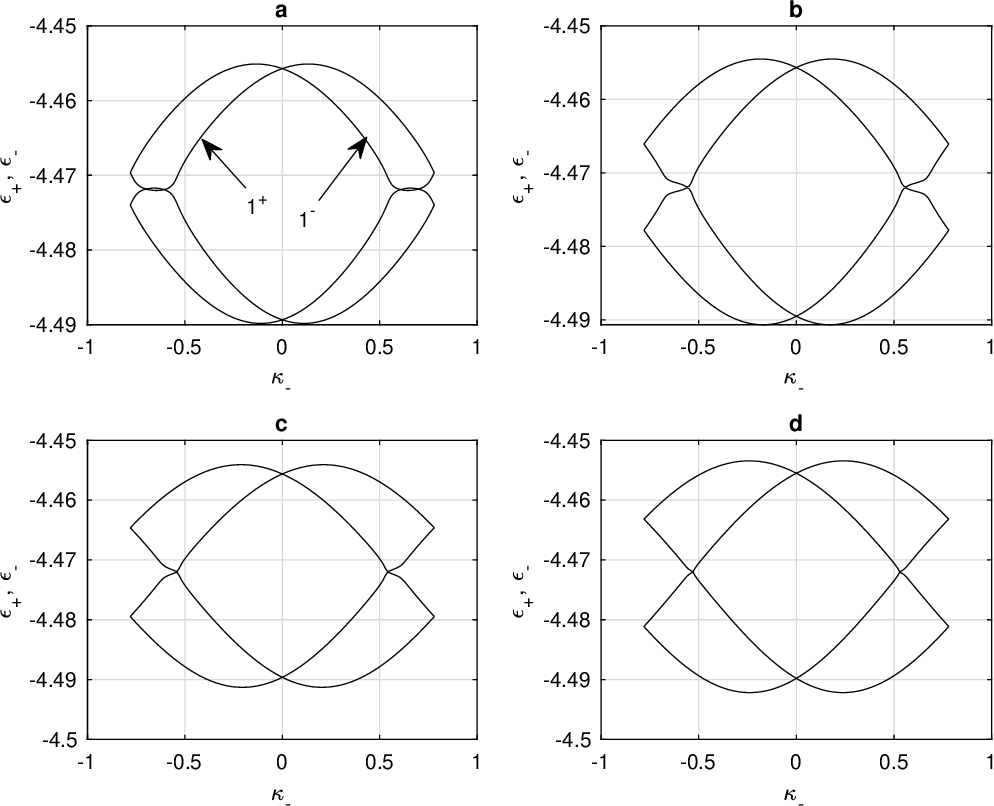}}
}
\caption{The energy bands $\Ge_+$ and  $\Ge_-$  versus $\Gk_-$  when $\Gk_+=-0.664\pi/c_+$ (curve $1^-$) and $\Gk_+=0.664\pi/c_+$ (curve $1^+$) 
for zigzag nanotubes $\CA(n,0)$ when $n=5$ (a),  $n=7$ (b),  $n=9$ (c), and $n=15$ (d).} 
\label{fig17}
\end{figure}

For numerical tests we employ the atomic units. Based on  the first nearest-neighbor tight-binding approximation we select the external potential to be 
\beq
\CV(\Br)=\sum_{\mu=A,B}\sum_{s=-L_n}^{L_n-1}\sum_{j=-L_m}^{L_m-1}{'}v^\mu_{sj}(\Br),
\label{6.35}
\eeq
where $v^\mu_{sj}(\Br')$, in the  local spherical coordinates associated with the atom $\mu=A,B$
of the $(s,j)$-th parallelogram of periods, is given by $v^\mu_{sj}(\Br') =-\fr{Ze^2}{r'}$ if $r'\le  a/2$ and 0 otherwise,
$Z=6$, and $e=1$. 

The 2d energy dispersion relations throughout the first Brillouin zone for two chiral nanotubes $\CA(5,2)$ and $\CA(10,2)$
are shown in Figs \ref{fig9}  and \ref{fig10}. These diagrams demonstrate that in both cases there is a curve in the Brillouin zone along which $\Ge_+=\Ge_-$.
The zero band gap property  is confirmed by a detailed study of the associated  1d dispersion relations (Figs \ref{fig11} and  \ref{fig12}). These diagrams indicate that  there exist two symmetric zero band gap curves (the second curve is not visible in  Figs \ref{fig9}  and \ref{fig10}).  For
the nanotube $\CA(5,2)$, when
 $ -0.64\pi/c_+\le \Gk_+\le  -0.57\pi/c_+$
and  $0.57\pi/c_+\le \Gk_+\le 0.64\pi/c_+$,  there is $\Gk_-\in(-\pi/c_-,\pi/c_-)$ such that  both eigenvalues coincide.
In the case of the tube $\CA(10,2)$  such curves are  shorter and observed for  $0.62\pi/c_+\le  |\Gk_+| \le 0.64\pi/c_+$ (in 
Fig. \ref{fig12} only the dispersion curves for negative $\Gk_+$ are shown).

The 2d energy dispersion relation for an armchair tube $\CA(10,10)$ in the first square Brillouin zone
is shown in Fig. \ref{fig13}. It is seen that the upper and lower surfaces form a Dirac cone (there is another one in the first Brillouin zone -
it is not visible in Fig. \ref{fig13}). The energy bands for sample armchair  tubes ($n=5,7, 10, 15$)  are given in Fig. \ref{fig14}.  In all the cases tested the minimum of the gap is attained at 
approximately $\Gk_+=\Gk_-=\pm 0.664 \pi/c_+$. When $n$ is growing, the curves in the neighborhood of
the points where $\Ge_+=\Ge_-$ become indistinguishable from the Dirac cone.
What also is observed is that for thinner armchair tubes, when $n\le 5$,
the upper and lower dispersion curves are smooth at the points where $\Ge_+=\Ge_-$. When $n$ increases ($6\le n\le 10$), the curves are drifting from each other, and a small gap between the curves become visible. 
The further increase of the armchair number eliminates the gap, and  two Dirac points in the Brillouin zone emerge (see Fig. \ref{fig14}). 

 Our computations demonstrate that both characteristics, the nanotube chirality and its radius, make impact on the electronic structure.
Sample one-dimensional energy dispersion relations for chiral nanotubes $\CA(15,m)$ are shown in Fig. \ref{fig15}. 
For smaller $m$, $1\le m\le 7$, the upper and lower curves touch each other at two symmetric points (Fig. \ref{fig15}a,b).  For $m=8, 9$, there is a small gap that does not vanish
at the points $c_+|\Gk_+|/\pi \in[0.655,0.665]$, where the smallest gap is observed. For $m=10,11,\ldots,15$, there is a zero gap at $\Gk_+=\pm 0.664\pi/c_+$;
the two curves form Dirac cones whose vertices become Dirac points.  
Our numerous tests, including those 
demonstrated in Fig. \ref{fig15}, do not indicate that there is any 
correlation between the property of a nanotube being metallic and the property of the number $n-m$ being a multiple of 3.

We also report our results on zigzag nanotubes $\CA(n,0)$. We have not found any $n$ for which even a small gap between the upper and lower dispersion curves
is observed. Fig. \ref{fig16} demonstrates the 2d band gap structure  for a narrow zigzag tube $\CA(6,0)$. To clarify the electronic structure of zigzag nanotubes, in Fig. \ref{fig17}  we show the energy bands for nanotubes
$\CA(5,0)$, $\CA(7,0)$, $\CA(9,0)$, and $\CA(15,0)$. Their diameters are 3.92 nm, 5.48 nm, 7.05 nm, and 11.75 nm, respectively. For small radius zigzag nanotubes (see Fig. \ref{fig17}a for $\CA(5,0)$),  when $\Gk_+=-0.664 \pi/c_+$, there
is not just a zero-gap-point $\Gk_-$ but a curve along which the upper and lower dispersion curves overlap.  It turns out that the overlap occurs for all $c_+|\Gk_+|/\pi\in(0.65,   0.68)$,
that is there are two zero band gap domains $\Go_\pm$ in the Brillouin zone, and these domains are 2d. The 2d zero gap domain is bigger for $n=4$ and smaller for $n=6$. This means that small radius zigzag nanotubes  $\CA(n,0)$ ($n=4,5,6$) are metallic. This is consistent with the 
 results obtained by the DFT method  \cite{bla},
 \cite{sha}  and a Coulomb interaction model \cite{barn}. For  nanotubes $\CA(n,0)$, $n=7,\ldots,13$,  our calculations show that 
the zero band gap domains are not a 2d domain anymore - they degenerate into two symmetric short curves.  For $n\ge 14$, the two curves touch each other at two symmetric points in the first Brillouin zone, form Dirac like cones, and 
the electronic structure of such  zigzag nanotubes is that of graphene.

Finally, we note that our calculations implemented on the basis of the second  nearest-neighbor tight-binding approximation scheme do not demonstrate any visible improvements
since additional integrals generated by the scheme are of the order $10^{-5}$ and comparable with the magnitude of the imaginary part of the eigenvalues
produced by first nearest-neighbor tight-binding approximation.

\setcounter{equation}{0}

\section{Conclusion}

To compute the electronic structure of SWCNTs by means of tight-binding approximation, we have introduced  a concept of the reciprocal tube and developed
 an analogue of the Bloch theory for armchair, zigzag, and chiral nanotubes. 
For describing  the real and reciprocal nanotubes we have employed cylindrical coordinates
in the real and reciprocal 3d spaces. The unit cell of the real and reciprocal tubes used is a single hexagon and parallelogram, respectively,  rolled around the tubes.
 The first Brillouin zone we have constructed  is a rectangle in the chiral case and a square for armchair and zigzag nanotubes.  Based on this description we have formulated  and proved 
the Bloch theorem for nanotubes. It states  that first, the Bloch states are  quasi-periodic, $\psi(\CT_j^\pm\Br; \Bk)=e^{\pm ij\Gk^\pm c_\pm}\psi(\Br; \Bk)$,
and second, they admit the representation $\psi(\Br; \Bk)=e^{\Br\cdot\Bk} u(\Br; \Bk)$.  Here, $u(\Br; \Bk)$ is invariant with respect to the translation-rotation transformations
$\CT^\pm_j$ and $\tilde\CT^\pm_j$ acting on the real and reciprocal tubes simultaneously, $u(\CT_j^\pm\Br; \tilde\CT_j^\pm\Bk)=u(\Br;\Bk)$.
It needs to be underlined that these statements are valid only on the surface of the nanotube, and the external potential is required to possess the translation-rotation symmetry on 
the surface only. That is why in the tight-binding approximation scheme applied to nanotubes, in addition to the standard approximations, we make another approximation by
assuming that
the external potential and the Bloch sums defined in  the 3d space satisfy the Bloch properties not only on the surface but also in its neighborhood.
The double Bloch sums used in the tight-binding approximation procedure are constructed
by employing two intersecting helices that results in macroscopically large numbers of terms in both sums and, as a consequence, a 2d first Brillouin zone.
It is a rectangle in the chiral case and squares in the armchair and zigzag cases. 

The electronic band structure calculations by means of the first nearest-neighbor tight-binding approximation we have implemented show the efficiency of the numerical method for all ranges of nanotubes, small, intermediate and large radius-nanotubes. Our results  demonstrate that small radius zigzag nanotubes $\CA(n,0)$ with $n=4,5,6$
are metallic: their upper and lower dispersion curves overlap, and moreover, there are two symmetric 2d subdomains in the Brillouin square where the dispersion surfaces overlap.
Zigzag nanotubes $\CA(n,0)$ with  $6 \le n \le 13$ are semi-metallic: there is no overlapping of the dispersion relation curves but there are two intervals for $\Gk_+$     
for which there is a point $\Gk_-$ such that the two curves touch each other. When $n\ge 14$, the two curves touch each other only at two points in the first Brillouin zone,
the curves form a Dirac cone, the curvature effect become insignificant, and the electronic structure of such zigzag nanotubes is that of graphene. 

We have not found any chiral nanotube for which our simulations reveal overlapping of the dispersion curves. For narrow chiral tubes, the upper and lower dispersion curves touch each other
along two symmetric curves $\Gg^\pm(\Gk_+,\Gk_-)$ lying in the first Brillouin zone (a  rectangle). When the nanotube radius increases, each  curve $\Gg^\pm$ degenerates into  a 
single point.  Its further increase leads to a small, visible however, gap between the curves. For instance, for chiral nanotube $\CA(15,m)$ with $n=1,\ldots,4$, we observe
an attachment of the dispersion curves, and its electronic structure is close to that of a zigzag tube $\CA(15,0)$.  For $n=5,\ldots, 10$, there is a small visible gap, while for $n=11, \ldots,14$, this gap decreases and becomes practically not visible.  
For example, there is no visible difference between the electronic structures of a chiral tube $\CA(15,14)$ and an armchair tube $\CA(15,15)$.

Among all SWCNTs the armchair nanotubes electronic structure is the closest to that of graphene. The upper and lower dispersion curves look like a cone in the area where they close to each other. However, for narrow ($n=3,4,5$) and intermediate-radius ($n=6,\ldots, 15$) tubes $\CA(n,n)$,  there are some differences.
Although  the curves related to narrow armchair tubes touch each other at only two symmetric points in the first Brillouin zone, the curve are smooth in the vicinity of these points. 
For $n=6, \ldots, 10$, there is a small band gap that decreases with the further increase of $n$. For $n\ge 11$,  the points where the curves attach become  Dirac points.

\vspace{.1in}

{\bf Data accessibility.} This article does not contain any additional data.

{\bf Declaration of AI use.} I have not used AI-assisted technologies in creating this article.

{\bf Authors’ contributions.} Y.A.A.: conceptualization, formal analysis, investigation, methodology, software,
validation, writing—original draft, writing—review and editing.

{\bf Conflict of interest declaration.} I declare I have no competing interests.

{\bf Funding.} This material is based upon work supported  by the Air Force Office of Scientific Research 
 under award number  FA9550-24-1-0177.


\begin{thebibliography}{99}


\bibitem{iij}  Iijima S. 1991 Helical microtubules of graphitic carbon.  \textit{Nature}  \textbf{354}, 56-58.

\bibitem{sai98}   Saito R, Dresselhaus G,  Dresselhaus MS. 1998 \textit{Physical Properties of Carbon Nanotubes}. London: Imperial College Press.
\bibitem{wal}   Wallace PR. 1947 The band theory of graphite.  \textit{Phys. Rev.} \textbf{71}, 622-634.

\bibitem{net}  Castro Neto AH, Guinea F, Peres NMR,
Novoselov KS,  Geim AK. 2009 The electronic properties of graphene.  \textit{Rev. Mod. Phys.} \textbf{81}, 109-162.

\bibitem{ham} Hamada N, Sawada S,  Oshiyama A. 1992 New one-dimensional conductors: graphitic microtubules.  \textit{Phys. Rev. Lett.} \textbf{68}, 1579-1581.

\bibitem{sai92a}  Saito R, Fujita M, Dresselhaus G,  Dresselhaus MS. 1992 Electronic structure of graphene tubules based on $C_{60}$.
 \textit{Phys. Rev. B} \textbf{46}, 1804-1811.

\bibitem{sai92b}  Saito R, Fujita M, Dresselhaus G,  Dresselhaus MS. 1992 Electronic structure of chiral graphene tubules.
\textit{Appl. Phys. Lett.} \textbf{60}, 2204-2206.

\bibitem{aji} Ajiki H, Ando T. 1993 Electoronic states of carbon nanotubes.  \textit{J. Phys. Soc. Jpn} \textbf{62}, 1255-1266.

\bibitem{sai93}   Saito R, Dresselhaus G,  Dresselhaus MS. 1993 Electronic structure of double‐layer graphene tubules. \textit{J. Appl. Phys.} \textbf{73}, 494–500.


\bibitem{jis}  Jishi RA, Inomata D, Nakao K, Dresselhaus MS,  Dresselhaus G. 1994 Electronic and lattice properties of carbon nanotubes.   \textit{J. Phys. Soc. Jpn} \textbf{63}, 2252-2260.

\bibitem{bar} Barros EB, Jorio A, Samsonidze GG, Capaz RB, Souza Filho AG,
Mendez Filho J, Dresselhaus G, Dresselhaus MS. 2006 Review on the
symmetry-related properties of carbon nanotubes.  \textit{Phys. Rep.} \textbf{431}, 261-302.

\bibitem{bla} Blase X, Benedict LX, Shirley EL, Louie SG. 1994 Hybridization effects and metallicity in small radius carbon nanotubes. 
\textit{Phys. Rev. Lett.} \textbf{72}, 1878-1881.

\bibitem{kle} Kleiner A,  Eggert S. 2004 Band gaps of primary metallic carbon nanotubes.  \textit{Phys. Rev. B} \textbf{63}, 073408.

\bibitem{pop} Popov VN. 2004 Curvature effects on the structural, electronic and optical properties of isolated single-walled carbon
nanotubes within a symmetry-adapted non-orthogonal tight-binding model. \textit{New J. Phys.} \textbf{6}, 17.

\bibitem{barn}  Barnett R, Demler E,  Kaxiras E. 2005 Electron-phonon interaction in ultrasmall-radius carbon nanotubes. 
 \textit{Phys. Rev. B} \textbf{71}, 035429.

\bibitem{yu}  Yu HM,  Banerjee AS. 2022 Density functional theory method for twisted geometries with
application to torsional deformations in group-IV nanotubes. \textit{J. Comp. Phys.} \textbf{456}, 111023.


\bibitem{aga} Agarwal S, Banerjee AS. 2024 Solution of the Schrödinger equation for quasi-one-dimensional
materials using helical waves. \textit{J. Comp. Phys.} \textbf{496},  112551.

\bibitem{blo} Bloch F. 1929 Über die Quantenmechanik der Elektronen in Kristallgittern.  \textit{Z.  f. Physik} \textbf{52}, 555-600.

\bibitem{kitt} Kittel C. 1963 \textit{Quantum Theory of Solids.} New York.: John Willy \& Sons, Inc. 

\bibitem{kax} Kaxiras E, Joannopoulos JD. 2019  \textit{Quantum Theory of Materials.} Cambridge: Cambridge University Press, 

\bibitem{whi}   White CT, Robertson DH,  Mintmire JW. 1993 Helical and rotational symmetries of nanoscale graphitic tubules.
 \textit{Phys. Rev. B} \textbf{47}, 5485-5488.
   
   
\bibitem{zha}  Zhang DB,  Hua M,  Dumitric\u{a} T. 2008 Stability of polycrystalline and wurtzite Si nanowires via symmetry-adapted
tight-binding objective molecular dynamics.  \textit{J. Chem. Phys.}  \textbf{128},  084104.


\bibitem{sha} Shan B,  Cho K. 2005
First principles study of work functions of single wall carbon nanotubes, \textit{Phys. Rev. Lett.} \textbf{94}, 236602.

\end{thebibliography}
\end{document}